\PassOptionsToPackage{dvipsnames}{xcolor}
\documentclass[twocolumn, switch]{article} % Method A for two-column formatting

\usepackage{preprint}

\usepackage{amsmath, amsthm, amssymb, amsfonts}

\usepackage[numbers,square]{natbib}
\usepackage[T1]{fontenc}	% use 8-bit T1 fonts
\usepackage[colorlinks = true,
           linkcolor = purple,
            urlcolor  = blue,
            citecolor = cyan,
            anchorcolor = black]{hyperref}	% Color links to references, figures, etc.
\usepackage{nicefrac}		% compact symbols for 1/2, etc.
\usepackage{microtype}		% microtypography
\usepackage{lineno}		% Line numbers
\usepackage{float}			% Allows for figures within multicol
\usepackage{newfloat}
\DeclareFloatingEnvironment[name={Supplementary Figure}]{suppfigure}
\usepackage{sidecap}
\sidecaptionvpos{figure}{c}

\usepackage{titlesec}
\titlespacing\section{0pt}{12pt plus 3pt minus 3pt}{1pt plus 1pt minus 1pt}
\titlespacing\subsection{0pt}{10pt plus 3pt minus 3pt}{1pt plus 1pt minus 1pt}
\titlespacing\subsubsection{0pt}{8pt plus 3pt minus 3pt}{1pt plus 1pt minus 1pt}

\usepackage[dvipsnames]{xcolor} % colors
\usepackage[toc,page]{appendix} % appendix
\usepackage{subcaption} %subfigures
\usepackage{booktabs} %tables
\usepackage{dsfont} % indicator function
\usepackage{pifont}% http://ctan.org/pkg/pifont
\usepackage{stmaryrd} % \shortrightarrow
\usepackage{makecell} % cell table
\usepackage{graphicx,multirow} % rotated row
\usepackage{colortbl} % color table
\usepackage{url}  % url
\usepackage[binary-units=true]{siunitx}
\usepackage{caption} % centering table cap.
\usepackage{float} % figure at top
\usepackage{threeparttable}
\usepackage{tabularx}

\definecolor{goldyellow}{RGB}{252, 214, 0}
\definecolor{redd}{RGB}{199, 53, 0}
\definecolor{greenn}{RGB}{213,232,212}
\definecolor{bluee}{RGB}{218,232,252}

\definecolor{Gray}{gray}{0.96}
\newcolumntype{g}{>{\columncolor{Gray}}c}
\usepackage{etoolbox}

\graphicspath{ {./figures/} }
\title{Valuing Vicinity: Memory attention framework for context-based semantic segmentation in histopathology}

\usepackage{authblk}

\author[1,2]{Oliver Ester}
\author[1,2]{Fabian Hörst}
\author[3]{Constantin Seibold}
\author[1]{Julius Keyl}
\author[4]{Saskia Ting}
\author[5]{Nikolaos Vasileiadis}
\author[5]{Jessica Schmitz}
\author[6]{Philipp Ivanyi}
\author[2,7]{Viktor Grünwald}
\author[5]{Jan Hinrich Bräsen}
\author[1,2]{Jan Egger}
\author[1,2,8,\thanks{\tt{jens.kleesiek@uk-essen.de}}]{Jens Kleesiek}
	
\affil[1]{Institute for AI in Medicine (IKIM), University Hospital Essen (AöR), Essen, Germany}
\affil[2]{Cancer Research Center Cologne Essen (CCCE), West German Cancer Center Essen, University Hospital Essen (AöR), Essen, Germany}
\affil[3]{Institute of Anthropomatics and Robotics, Karlsruhe Institute of Technology (KIT), Karlsruhe, Germany}
\affil[4]{Institute of Pathology, University Hospital Essen (AöR), University of Duisburg-Essen, Essen, Germany}
\affil[5]{Nephropathology Unit, Institute for Pathology, Hannover Medical School, Hannover, Germany}
\affil[6]{Department of Hematology, Hemostasis, Oncology and Stem Cell Transplantation, Hannover Medical School, Hannover, Germany}
\affil[7]{Clinic for Medical Oncology, Clinic for Urology, West German Cancer Center, University Hospital Essen (AöR), Essen, Germany}
\affil[8]{German Cancer Consortium (DKTK), Partner Site Essen, Germany}

\begin{document}

\twocolumn[ % Method A for two-column formatting
  \begin{@twocolumnfalse} % Method A for two-column formatting
  
\maketitle

\begin{abstract}
	%%%
	The segmentation of histopathological whole slide images into tumourous and non-tumourous types of tissue is a challenging task that requires the consideration of both local and global spatial contexts to classify tumourous regions precisely. The identification of subtypes of tumour tissue complicates the issue as the sharpness of separation decreases and the pathologist's reasoning is even more guided by spatial context. However, the identification of detailed types of tissue is crucial for providing personalized cancer therapies.
		Due to the high resolution of whole slide images, existing semantic segmentation methods, restricted to isolated image sections, are incapable of processing context information beyond.
		To take a step towards better context comprehension, we propose a patch neighbour attention mechanism to query the neighbouring tissue context from a patch embedding memory bank and infuse context embeddings into bottleneck hidden feature maps. Our memory attention framework (MAF) mimics a pathologist's annotation procedure -- zooming out and considering surrounding tissue context. The framework can be integrated into any encoder-decoder segmentation method. We evaluate the MAF on a public breast cancer and an internal kidney cancer data set using famous segmentation models (U-Net, DeeplabV3) and demonstrate the superiority over other context-integrating algorithms -- achieving a substantial improvement of up to $17\%$ on Dice score. The code is publicly available at \url{https://github.com/tio-ikim/valuing-vicinity}
	%%%%
\end{abstract}

%\keywords{First keyword \and Second keyword \and More} % (optional)
\vspace{0.25cm}

  \end{@twocolumnfalse} % Method A for two-column formatting
] % Method A for two-column formatting

%\begin{multicols}{2} % Method B for two-column formatting (doesn't play well with line numbers), comment out if using method A

%%%%%%%%%%%%%%%  Main text   %%%%%%%%%%%%%%%
% \linenumbers

\section{Introduction}
In the digital age of histopathology, specialized scanners digitize a tissue specimen with suspected cancer into an image at high magnification, resulting in a whole slide image (WSI). Tumourous tissue can be identified, graded and the most promising therapy recommended. 
The response of therapies is yet not fully understood and more research about the tumour microenvironment (TME) is ongoing \cite{junttila2013influence}. One line of research is the detailed cell analysis of subtypes of tissue. In this work, we focus on the identification of detailed types of tissue first.

\begin{figure*}[!t]
	\centering
	\includegraphics[width=0.9\textwidth]{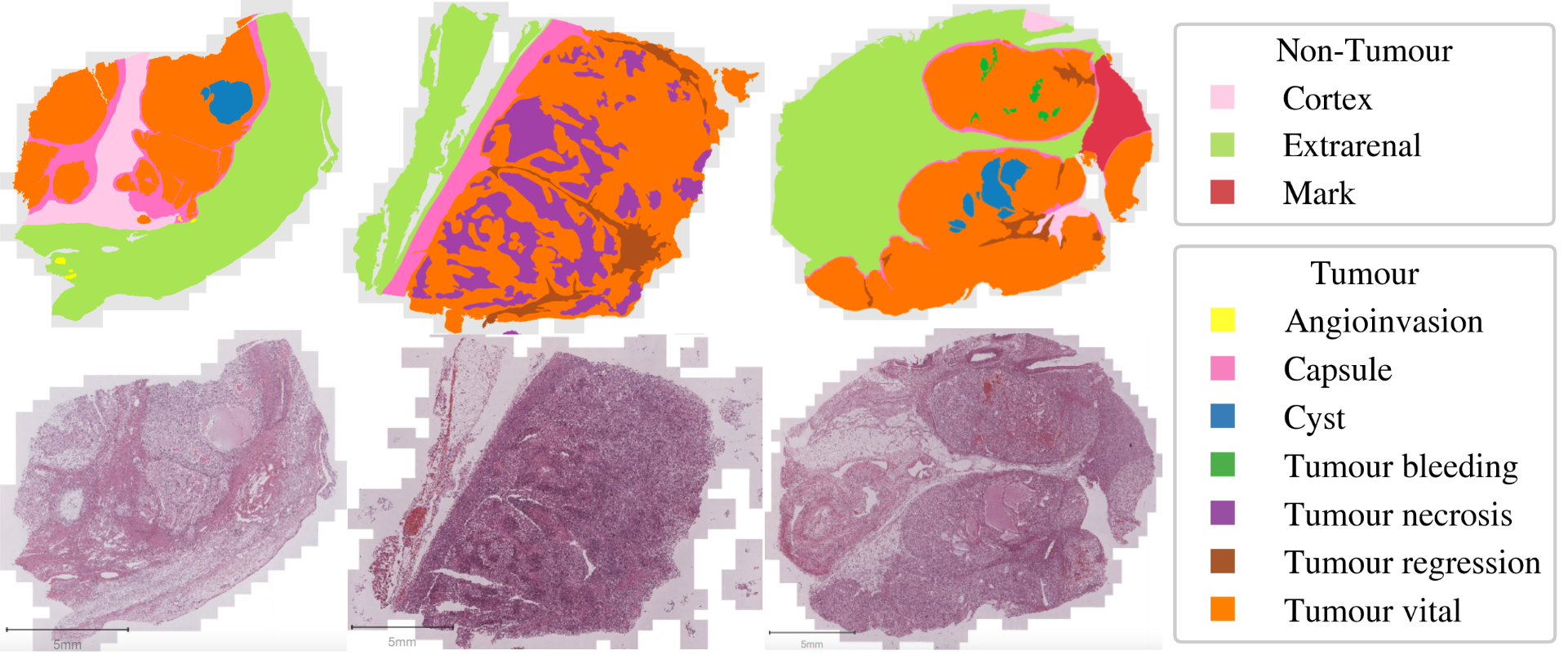}
	\caption{RCC examples -- Top: Annotations of tumourous and non-tumourous subtypes. Bottom: H\&E WSIs with kidney cancer.}
	\label{wsi}
\end{figure*}

Figure \ref{wsi} shows an example hematoxylin and eosin (H\&E) WSI with renal cell cancer (RCC) and its corresponding extensive annotations of subtypes of tissue. To identify a specific type of tissue, a pathologist examines a slide section at high magnification and then considers neighbouring tissue to integrate context information into the decision. As the manual annotation process is a tedious, complex task, there is ongoing research for developing WSI segmentation algorithms. The most promising algorithms for WSI segmentation are based on supervised, parameterizable convolutional neural networks (CNN) that are trained on a set of WSIs and their ground truth annotations.
However, since a single WSI at the highest resolution exceeds the hardware limits of modern GPUs, most methods decompose it into a set of patches while trying to find a balance between a narrower field of view (FOV) with less context information and a lower physical image resolution with fewer tissue details \cite{wang2019pathology}.
Either way, the algorithm is withheld with crucial information.

\vspace{-0.1cm}
In this work, we present a novel memory extension framework for CNN encoder-decoder architectures in semantic segmentation to improve the exploitation of context information around WSI patches.
\vspace{-0.3cm}

\subsection{Related work}

\vspace{-0.1cm}
Semantic segmentation is the computer vision task to assign each pixel to its corresponding class. Due to the dense prediction characteristic and often cohesive label regions (annotations), specialized segmentation CNNs -- like U-Net \cite{ronneberger2015unet} or DeepLabV3 \cite{chen2017deeplab} -- showed great performance on various segmentation tasks.

\vspace{-0.15cm}
\textit{Semantic segmentation of WSIs.} The segmentation of WSIs (e.g. to detect specific tissue classes) poses a considerable challenge:
Due to the huge resolution of WSIs (larger than 150k $\times$ 150k px per WSI), applying existing state-of-the-art (SOTA) segmentation algorithms directly onto the slide is impeded by current GPU hardware limitations.
Therefore, in the infancy of semantic segmentation of WSIs, segmentation algorithms were applied in a patch-wise manner \cite{wang2019pathology}. 
Yet, the patch processing is not trivial and \citet{jin2020foveation} showed that the segmentation quality heavily depends on the selection of two patch hyperparameters -- downsampling (resolution) and FOV (spatial extent of context). To overcome the selection process, they developed a \textit{foveation module}, which learns to dynamically select the best trade-off between both hyperparameters.

\textit{Context integration.}
Other works exploit context information to enlarge the FOV while simultaneously preserving a detailed resolution. To do so, concentric context patches with lower physical resolution around a central patch are additionally fed into a CNN and context information is merged \cite{class_breast,tokunaga2019adaptive,gu2018multiresolution,li2018multiscale,autom_segm}. 
\cite{schmitz2021} developed different fusion architectures -- based on the U-Net -- with multiple encoders for patches with different FOVs and fused hidden features in the bottleneck. However, the success of the context fusion heavily depends on the context FOV and the type of tumour.
At the same time, \cite{vanrijthoven2020hooknet} developed the HookNet (two parallel U-Net architectures with a context and a target branch) that hooks a wider FOV into the target branch by spatially aligning the feature map crop from the context branch to the target feature map.
%\vspace{-0.5cm}
\textit{Segmentation and attention mechanism.}
After the successful adaption of the Transformer architecture \cite{vaswani2017attention} into the computer vision field \cite{dosovitskiy2021image}, many ideas about exploiting the Transformer architecture for semantic segmentation were developed \cite{wang2021mixed,strudel2021segmenter,li2021medical,chen2021transunet,xie2021segformer,zheng2021rethinking}. Mostly, these methods integrate Transformer modules into encoder-decoder architectures to increase the receptive field using attention mechanisms.
\cite{chen2021transunet} added a Transformer into the U-Net bottleneck. Thereby, the self-attention mechanism can attend to all (spatial) features of the feature map.
\cite{guo2021selfattention} introduced an external attention mechanism to exploit potential correlation with other samples by attending an embedding memory.
Based on the external attention mechanism, \cite{wang2021mixed} proposed a Mixed Transformer Module that extends the encoder and decoder depths. Their module consists of three attention mechanisms: a local, a global, and an external.
\nopagebreak The local one attends the close context by a local window constraint -- the global one attends tokens by \nopagebreak a row and column constraint globally -- the external one attends a memory of queries and keys of the entire data set.

\subsection{Our contribution}

We were inspired by the external memory and developed a memory for spatial representations of each WSI by storing patch embeddings. Related to \cite{chen2021transunet}, we integrate an attention mechanism in the bottleneck of an encoder-decoder architecture. For the interaction between the external memory and the attention mechanism, our architecture attends neighbouring patch embeddings stored in the external memory while segmenting a given patch. By doing so, our patch neighbour \textbf{M}emory \textbf{A}ttention \textbf{F}ramework (MAF) enriches out-of-sample context information into the patch segmentation. The attention module learns to dynamically select relevant context information. We compare the MAF's capability of context integration to the context-integrating msY-model of \cite{schmitz2021} on two cancer data sets: renal cell cancer and breast cancer.

We summarize our contributions as:
\begin{enumerate}
	\item We propose a novel out-of-sample attention-based extension for arbitrary encoder-decoder architectures and a corresponding memory framework to integrate neighbourhood context information from a memory into the patch segmentation process and 
	\item demonstrate that our method can be beneficial to tissue segmentation of kidney cancer and tumour segmentation of breast cancer.
	\item We show that our attention mechanism is able to shed light on the impact of context information in terms of model explainability.	      
\end{enumerate} 

\begin{figure*}[ht!]
	\centering
	\includegraphics[width=0.7\textwidth]{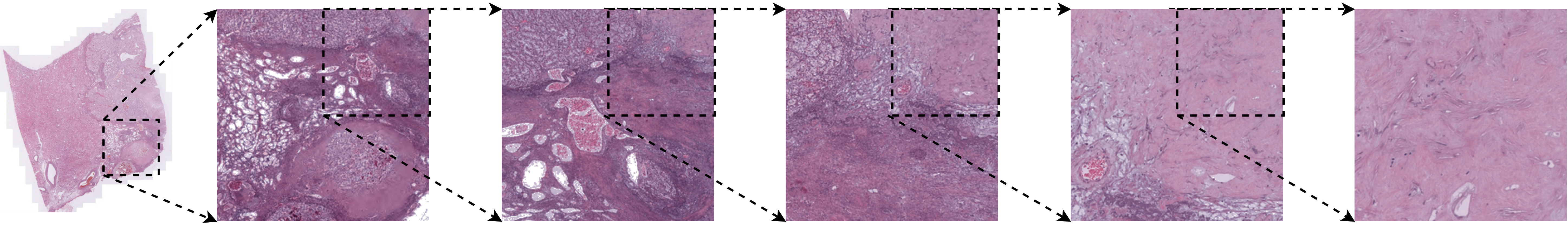}
	\caption{Example of patches with $256 \times 256$ px and different pixel resolutions ($\rightarrow$ different FOVs). From left to right: Thumbnail, \SI{22.14}{\micro\metre}, \SI{11.07}{\micro\metre}, \SI{5.53}{\micro\metre}, \SI{2.77}{\micro\metre}, \SI{1.38}{\micro\metre}}
	\label{patch_zoom}
\end{figure*}

\begin{figure*}[ht!]
	\centering
	\includegraphics[width=1\textwidth]{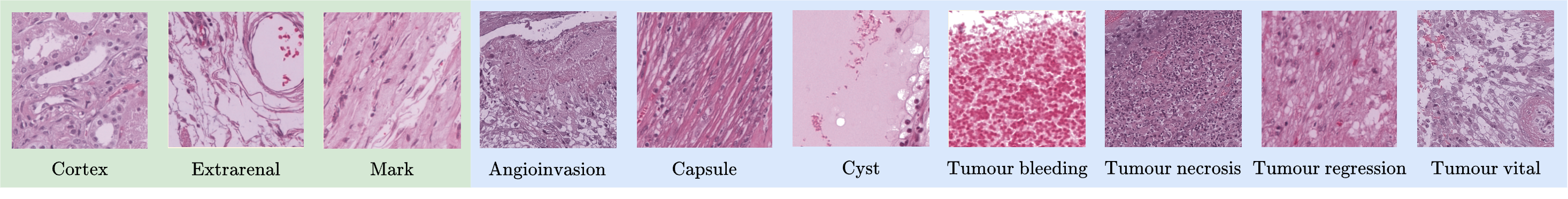}
	\caption{Patch example for each type of tissue with an edge length of 256 px and \SI{353,28}{\micro\metre} (\SI{1.38}{\micro\metre}/px). \textcolor{greenn}{$\blacksquare$} Non-tumourous \textcolor{bluee}{$\blacksquare$} Tumourous}
	\label{patch_classses}
\end{figure*}

\begin{figure}[!t]
        \centering
	\includegraphics[scale=.3]{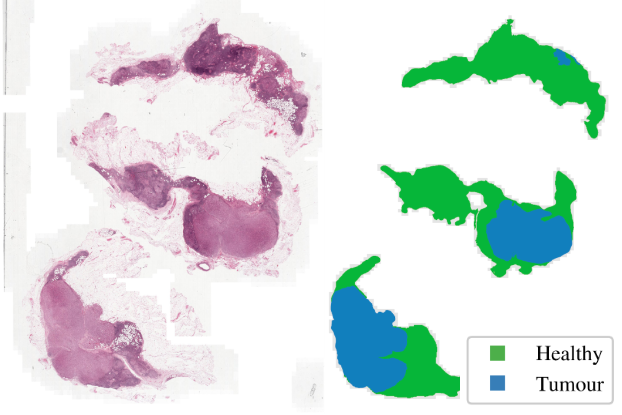}
	\caption{CY16 example -- Left: H\&E WSI with lymph node macrometastasis of breast cancer. Right: Annotations of healthy and tumourous tissue.}
	\label{cy16}
\end{figure}

\section{Materials}

We benchmark the MAF on an internal kidney cancer data set with renal cell carcinoma (RCC) and exhaustive annotations of various tumourous and non-tumourous types of tissue but also show the contribution of the MAF on a subset of the breast cancer data set from the CAMELYON16 (CY16) challenge \cite{camelyon16} following \cite{schmitz2021}.

\subsection{RCC} 
\label{rcc_ds}
The RCC data set consists of 175 WSIs of patients with metastatic renal
cell carcinoma undergoing nephrectomy.\footnote[1]{Approved by local ethics board: 8682\_BO\_K\_2019, 10183\_BO\_K\_2022} One representative archival paraffin block of the tumour was selected and sections of all specimens were stained with H\&E using routine procedures.
The slides were digitized at a pixel resolution of \SI{0.1729}{\micro\metre}.
We show a tissue example at different pixel resolutions in Fig. \ref{patch_zoom}.

\vspace{+3pt}
\noindent\textbf{Annotation}
The annotation of 10 different types of tissue was performed using QuPath \cite{qupath} under the close supervision of a trained nephropathologist. First, the tissue area was detected by thresholding followed by manual verifications. Subsequently, the tissue mask was manually separated into either tumourous or non-tumourous types: The tumourous regions were subclassified into either \textit{Tumour vital}, \textit{Tumour regression}, \textit{Tumour necrosis}, \textit{Tumour bleeding}, \textit{Angioinvasion}, \textit{Capsule} or \textit{Cyst} (resulting in 7 types of tumour tissue), the non-tumourous regions into either \textit{Extrarenal}, \textit{Cortex} or \textit{Mark} (resulting in 3 types of non-tumour tissue).
Fig. \ref{patch_classses} shows patch examples of size $256 \times 256$ px of each type of tissue at a pixel resolution of \SI{1.38}{\micro\metre} -- resulting in an edge length of \SI{352,72}{\micro\metre}.

\subsection{CY16} 
\label{cy16_ds}
We also validate our results on a publicly available data set of the CAMELYON16 challenge. The Cancer Metastases in Lymph Nodes (CAMELYON) 2016 challenge \cite{camelyon16} provided WSIs of sentinel lymph nodes with and without metastases of breast cancer. Following \cite{schmitz2021}, we select the same 20 WSIs with at least one lymph node macrometastasis. The slides were digitized at a pixel resolution of \SI{0.243}{\micro\metre}. In the following, we refer to the subset of CAMELYON16 WSIs as CY16.
Note, that the complexity of detecting tumourous tissue in the CY16 data set is assumed to be lower than the complexity of identifying all subtypes of tissue in the RCC data set due to the RCC's high histopathologic heterogeneity \cite{cai2020ontological}.

\vspace{+3pt}
\noindent\textbf{Annotation}
The data set provides annotations of the metastatic tissue. To come up with a tissue annotation, we followed \cite{schmitz2021} and applied thresholding using QuPath and created a \textit{Healthy} tissue class by subtracting the \textit{Tumour} annotation from the tissue annotation -- resulting in 2 tissue types -- \textit{Tumour} and \textit{Healthy}. Fig. \ref{cy16} shows an example WSI with its corresponding annotations.

\subsection{Patch Extraction}

For both data sets, we split each WSI into non-overlapping patches using OpenSlide \cite{openslide}, resized their original pixel resolution (downsampling) and applied the Macenko-normalization \cite{macenko}. We omitted all patches with no annotation overlap ($\rightarrow$ background).

To analyse the impact of physical resolution versus FOV, we created multiple patch sets using different downsampling factors ($\mathit{ds}$)
%(RCC: 8, 16, 32, 64 -- CY16: 1, 2, 4, 8, 16)
, all of pixel size $256 \times 256 \times 3$ in RGB colour space. A larger $\mathit{ds}$ corresponds to a larger FOV and a lower pixel resolution. Concurrently, for comparison with context-integrating models, context patches concentric to the central patch with identical pixel size but a larger FOV were extracted (related to \cite{schmitz2021}.
For the targets, we proceeded the same with the annotation masks and additionally enriched the patch metadata with the percentage class ratio, used as helper target.

\section{Methods}

In this section, we propose the \textbf{M}emory \textbf{A}ttention \textbf{F}ramework (MAF) for semantic segmentation of WSIs after first contextualizing the domain with preliminary definitions. 

\subsection{Preliminaries}
Let $w \in \mathbb{R}^{M \times N \times 3}$ be a WSI $w$ with $(M, N)$ spatial dimensions and three colour channels. Each $w$ is divided into a set of quadratic, non-overlapping patches $\mathcal{P}=\{p_{i,j}\}$, $p_{i,j} \in \mathbb{R}^{S \times S \times 3}$, where $i$ denotes the column position and $j$ the row position in a uniform, two-dimensional grid with the dimensions $n_\mathrm{x}$ and $n_\mathrm{y}$. The set of all WSI in the training set is denoted with $\mathcal{W} = \{w\}^{N_\mathrm{train}}$. In the case of multi-label segmentation with $C$ classes, the label for a patch $p_{i,j}$ is defined as the segmentation mask $ y_{i,j}^{seg} \in \{0,1\}^{S \times S \times C}$. Additionally, we define a second label ${y}_{i,j}^{cls} \in \mathbb{R}^{C}$ as the class distribution of all pixels in a patch.
In common segmentation methods \cite{fan2020ma,zhao2017pyramid,ronneberger2015unet,chen2017deeplab}, an encoder $f_\mathrm{enc}$ first maps a patch $p_{i,j}$ into a set of feature maps $\mathcal{F} = \{\mathit{feat}_0, \dots, \mathit{feat}_{l-1}, \mathit{feat}_l\}$ at different depth levels $l$. A decoder $f_\mathrm{dec}$ then learns to map $\mathcal{F}$ to the segmentation prediction $ \hat{y}_{i,j}^{seg}$.

\subsection{Memory attention framework}
\begin{figure*}[!t]
	\centering
	\includegraphics[width=\textwidth]{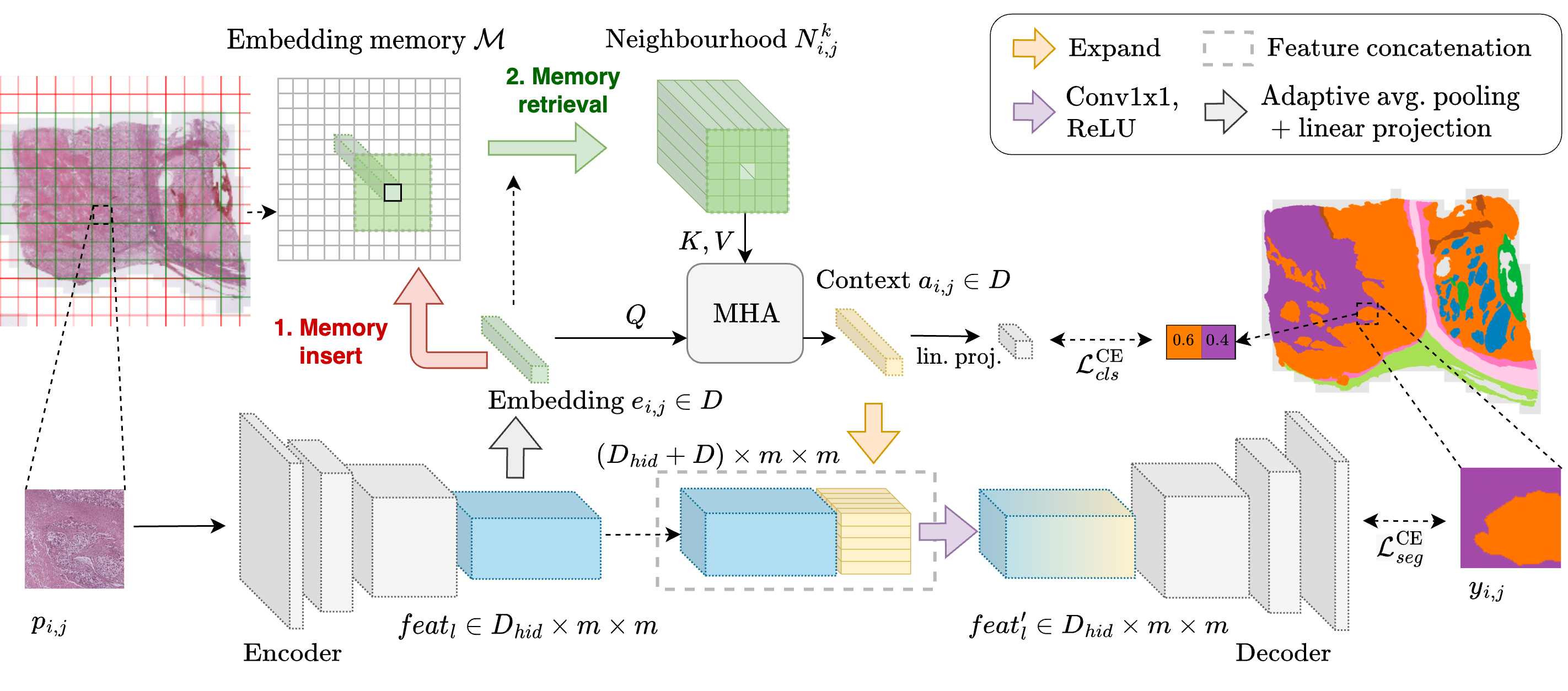}
	\caption{\label{att_memory} Overview of neighbourhood memory attention framework (MAF):
		An encoder-decoder architecture is extended by a patch memory attention mechanism to fuse context information into the segmentation process.
		1. The memory is built up in a patch-wise encoder run by compressing each patch through a gradient-free encoder-run (\textit{Memory insert}). 2. For each patch, the neighbourhood is retrieved from the embedding memory (\textit{Memory retrieval}), attended and the context embedding merged into the decoder run. A helper loss supports the context learning by predicting the patch's class ratios from the context embedding. For illustration purposes, we use a neighbourhood radius $k$ of 2.}
\end{figure*}

Fig. \ref{att_memory} shows an overview of our MAF based on a common encoder-decoder architecture but extended by a patch memory and an attention mechanism.
In our architecture, each patch is segmented individually but the framework enables an information flow of neighbourhood patches into each patch.
The neighbourhood memory attention mechanism exploits the deepest feature maps $\mathit{feat}_l$ from the encoder and creates a context-modified version $\mathit{feat}_l'$ which updates the set of feature maps $\mathcal{F}$ to $\mathcal{F}' = \{\mathit{feat}_0,\dots, \mathit{feat}_{l-1}, \mathit{feat}_l'\}$ before being decoded into $ y_{i,j}$. $\mathcal{F}'$ now is aware of context information.

We introduce different concepts that build up our MAF: 

\vspace{+3pt}
\noindent\textbf{Patch embedding}
To store every patch of a WSI $w$ in reasonable memory size, we learn compressed patch representations. 
For that, each patch $p_{i,j}$ is fed into the encoder $f_{enc}$ and a learnable compression function $f_{emb}$ (adaptive average pooling + linear projection) maps the features $\mathit{feat}_l \in \mathbb{R}^{D_{hid} \times m \times m}$ into an embedding $e_{i,j} \in \mathbb{R}^{D}$.

\begin{figure}[!t]
	\centering
	\includegraphics[width=0.45\textwidth]{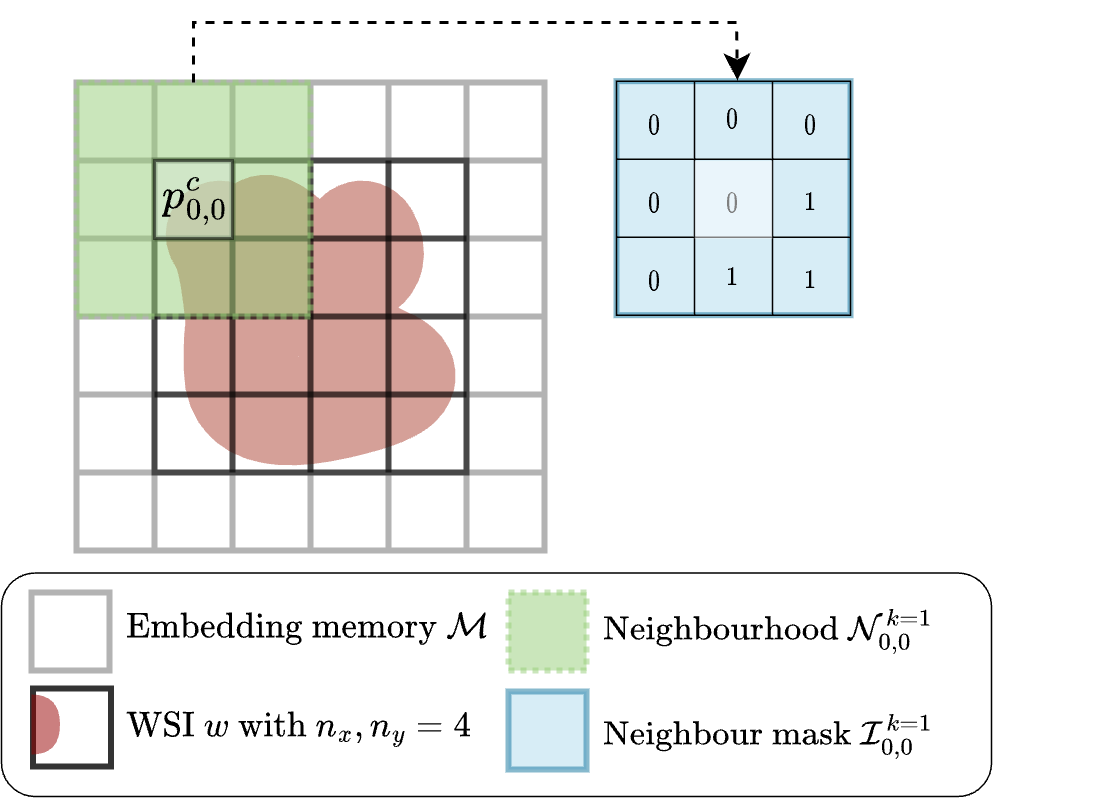}
	\caption{\label{memory}
		An example of a patched WSI $w$ with patch dimensions $n_x \times n_y = 4 \times 4$, a neighbourhood radius $k=1$ and the memory $\mathcal{M}$ with dimensions $(n_x + 2k) \times (n_y + 2k) \times D = (4 + 2 \cdot 1) \times (4 + 2 \cdot 1) \times D = 6 \times 6 \times D$. For the patch $p^c_{0,0}$ and its neighbourhood $\mathcal{N}^{k=1}_{0,0}$, we can derive the neighbour mask $\mathcal{I}^{k=1}_{0,0}$. The centreof the neighbour mask is always 0 by definition.}
\end{figure}

\vspace{+3pt}
\noindent\textbf{Embedding memory}
\label{subsec_mem}
A spatial embedding memory $\mathcal{M} \in \mathbb{R}^{(n_{\textrm{x}}+2k) \times (n_{\textrm{y}}+2k) \times D}$ for a WSI $w$ stores all patch embeddings $e_{i,j}$. At the boundary, we add a padding of size $k$ to the memory dimensions for neighbourhoods exceeding the WSI region (see Fig. \ref{memory}). 

\textit{1. Memory insert:} The memory $\mathcal{M}$ first has to be filled up, for each WSI $w$, by applying $f_{enc}$ and $f_{emb}$ to each patch $p_{i,j}$ and inserting the resulting compressed embeddings $e_{i,j}$ in $\mathcal{M}$. We update the memory once at the start of each epoch in a gradient-free forward pass balancing out-dated embeddings with resource-expensive updates.

\textit{2. Memory retrieval:} After the completion of the memory fill-up, we can retrieve the embeddings of the patch neighbourhood for a complete forward pass. This two-step forward pass is inevitable for the training and also test phase but marginally affects the run-time due to the first gradient-free forward pass allowing for larger batch size.
The backward pass does not affect the memory $\mathcal{M}$ instantly but merely updates the encoder weights and eventually leads to a delayed, epoch-wise update of the memory embeddings.

We also experimented with an \textit{online} memory retrieval (memory is not filled beforehand but the corresponding neighbourhood embeddings are determined in one forward pass) which suffers from exploding repetitions of encoder runs and exploding memory consumption with increasing neighbourhood size while, advantageously, offers \textit{up-to-date} embeddings due to the direct effect of the backpropagation. However, we could not find any model performance increase.

\vspace{+3pt}
\noindent\textbf{Patch neighbourhood}
\label{subsec_ptc}
The neighbourhood $\mathcal{N}$ defines the context information that can be reached by a patch $p$. We hypothesize that a larger neighbourhood provides more context information and therefore should improve the segmentation quality of $p$.
We define a concentric patch neighbourhood in the embedding space as
$\mathcal{N}^{k}_{i,j} = \{e_{i',j'}|i' \in [i-k, ..., i+k],j' \in [j-k, ..., j+k]\}$, $\mathcal{N}^{k}_{i,j} \subseteq \mathcal{M}$ for the central patch $p^{c}_{i,j}$, which is subject to the neighbourhood radius $k$.

To handle non-existing neighbour patches (e.g. at boundary areas or background patches), we also define a binary neighbour mask $\mathcal{I}^{k}_{i,j} = \{\mathds{1}_\mathcal{M}(i',j') | i' \in [i-k, ..., i+k],j' \in [j-k, ..., j+k]\}$ with 
\begin{equation}
	\mathds{1}_\mathcal{M}(i',j') = 
	\begin{cases}
		0\hspace{0.5cm} \text{if } \nexists e_{i',j'} \text{ or } (i',j') = (i,j) \\
		1\hspace{0.5cm} \text{else,}
	\end{cases}
\end{equation}
where the mask is $1$ if a patch embedding $e_{i',j'}$ exists in memory $\mathcal{M}$. We set $\mathds{1}_\mathcal{M}(i,j)$ constant to $0$ to exclude $p_{i,j}$ from its own neighbourhood and thereby avoid self-attention (see Fig. \ref{memory}).

\begin{figure}[!t]
	\includegraphics[width=0.5\textwidth]{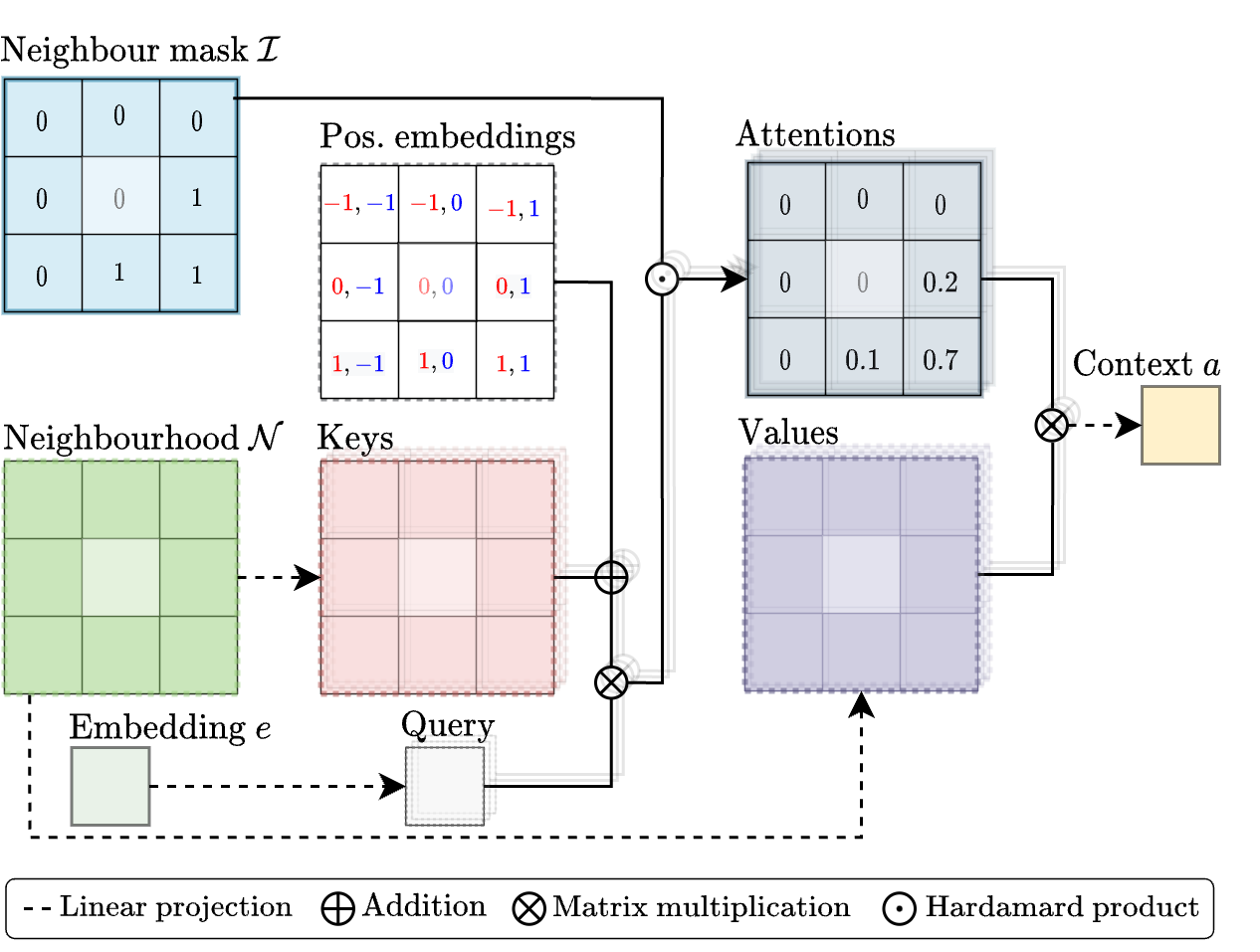}
	\caption{\label{attention} An example of the MHA mechanism: The embedding $e$ is linearly projected to the query, the neighbourhood $\mathcal{N}$ to keys and values, for each head respectively. We determine the attention scores by applying softmax to the matrix multiplication of the query with position-aware keys -- considering valid keys only by multiplying with the neighbour mask. Finally, we weight the values by the attention scores to receive the context embedding $a$. For illustration purposes, we use $k=1$.}
\end{figure}

\vspace{+3pt}
\noindent\textbf{Neighbourhood attention}
\label{subsec_att}
We hypothesize that the relevance of context information varies over space and type of tissue and thus should be learnable:
Hence, we make use of a Multi-Head-Attention (\textit{MHA}) module (adjusted from \cite{vaswani2017attention}) to enable the central patch $p^c_{i,j}$ in form of its embedding $e^c_{i,j}$ to query its neighbourhood $\mathcal{N}^{k}_{i,j}$ and obtain a context embedding $a_{i,j} \in \mathbb{R}^{D}$ (see example in Fig. \ref{attention}). The module learns to select and aggregate relevant context information providing a patch embedding as input:

For each head $h$, we project $e_{i,j}$ to the query vector $q_h$ and $\mathcal{N}^{k}_{i,j}$ to the keys and values matrices $K_h$ and $V_h$ and define the attention function as:

\begin{equation}
	\text{Attention}(q,K,V) = \text{softmax}(\mathcal{I} \frac{qK^T}{\sqrt{d}})V \text{ ,}
\end{equation}

where we mask the logits with our neighbour mask $\mathcal{I}$.
We alter the keys to $K_h'$ by adding position embeddings to $K_h$ (see Paragraph \ref{pos_emb}) and calculate the context embedding $a$ as:

\begin{equation}
	a = \text{Proj}(\underset{1..h}{\text{Concat}}(\text{Attention}(q_h,K'_h,V_h)))
\end{equation}

Throughout all experiments, we use $h=8$ and $d=128$ as the hidden dimension for $K$, $V$ and $q$.
In practice, we compute the MHA on a set of queries, keys, and values simultaneously.
Note, that the embedding $e_{i,j}$ origins from the second step of the forward pass and passes the gradients to the encoder while $\mathcal{N}^{k}_{i,j}$, being retrieved from the memory $\mathcal{M}$, is gradient-free.

\vspace{+3pt}
\noindent\textbf{Positional encoding}
\label{pos_emb}
The MHA enables the central patch $p^c$ to attend patches in the neighbourhood. However, in its raw form, the token order in $K$ and $V$ is permutation invariant and thus lacks any spatial awareness. We therefore add position embeddings to the keys. Following \cite{ramachandran2019stand}, we introduce learnable 2D position embeddings $B = \{b_{x,y} | x \in [-k,k], y \in [-k,k]\}$ with $b \in \mathbb{R}^d$ and relative patch coordinates $x$, $y$ to the central patch $p^c$ (see example in Fig. \ref{attention}). Each embedding $b$ is a concatenation of a row-offset representation and a column-offset representation:

\begin{equation}
	b_{x,y} = \text{Concat}(b^{row}_x, b^{col}_y) \text{ ,}
\end{equation}

with $b^{row} \in \mathbb{R}^{1/2d}$ and $b^{col} \in \mathbb{R}^{1/2d}$ being learnable parameters. We add $B$ to $K$ and receive position-aware keys $K'$. The position encodings are shared over the heads.

We also experimented with common fixed sinusoidal embeddings \cite{vaswani2017attention} and no position embeddings (see Table \ref{mod_var}).

\vspace{+3pt}
\noindent\textbf{Context fusion}
Eventually, a function $f_{fuse}$ fuses the context embedding $a_{i,j}$ into the feature maps $\mathit{feat}_l \in \mathbb{R}^{D_{hid} \times m \times m}$, resulting in $\mathit{feat}_l'$. It expands (copies) $a_{i,j} \in \mathbb{R}^D$ over the dimensions $m \times m$, concatenates with $\mathit{feat}_l$ and applies a $1 \times 1$ convolution over the dimensions $D_{hid} + D$ with output dimension $D_{hid}$. $\mathcal{F}$ is updated to $\mathcal{F}' = \{\mathit{feat}_0,\dots, \mathit{feat}_{l-1},{\mathit{feat}_l'}\}$
and $f_{dec}$ then predicts the segmentation mask $\hat{y}_{i,j}$ from $\mathcal{F}'$.
We also experimented with different convolution kernel sizes but did not observe any significant differences.
\vspace{+3pt}
\noindent\textbf{Targets and losses} 
The patch segmentation predictions $\hat{y}_{seg}$ are optimized on the common cross entropy loss $\mathcal{L}^{CE}_{seg}(\hat{y}_{seg}, y_{seg})$ given the patch segmentation labels $y_{seg}$.
Inspired by the Y-Net of \cite{mehta2018net}, we introduce a helper classification loss $\mathcal{L}_{cls}^{CE}(\hat{y}_{cls}, y_{cls})$ that aims to optimize the class distribution prediction $\hat{y}_{cls}$ given the true class distribution $y_{cls}$ from the patch metadata.
The class distribution prediction results from a linear projection of the context embedding $a$ -- the output of the MHA module. Note, that the central patch is always excluded from the neighbourhood -- so we hypothesize that predicting the central patch class distribution given the neighbourhood context only, can guide the loss in the MHA. We combine both losses using a weight $\lambda$ as $\mathcal{L}^{+} = (1-\lambda) \cdot \mathcal{L}_{seg}^{CE} + \lambda  \cdot \mathcal{L}_{cls}^{CE}$
and will refer to $\text{MAF}^{+}$ when optimizing on combined losses.

\section{Experiments}

\subsection{Experimental Setup}

\vspace{+3pt}
\noindent\textbf{Baseline and methods} 
We evaluate our method on two histopathological data sets, a breast cancer data set (see CY16 \ref{cy16_ds}) and a kidney cancer data set (see RCC \ref{rcc_ds}), and employ two baselines using patch-based segmentation architectures, namely U-Net \cite{ronneberger2015unet} and DeepLabV3 \cite{chen2017deeplab}. To show the general applicability of our new method, we extend both encoder-decoder architectures with the MAF and compare them with the patch-based baselines.
Finally, we compete our approach against another context-integrating architecture, the msY-Net \cite{schmitz2021} which learns from an additional context patch with larger FOV.

\vspace{+3pt}
\noindent\textit{Patch-based baseline:}
We first set up patch-based segmentation baselines to later measure the isolated effect of context contribution. Since \textit{the best} FOV for a patch remains unclear, we first evaluate the baseline models on different $\mathit{ds}$ factors -- keeping the patch dimension constant at $256 \times 256$ px.
Also, we test different encoder backends -- namely ResNet-50 and ResNet-18.

\vspace{+3pt}
\noindent\textit{Context benchmark:}
We then set up a SOTA context-integrating benchmark using the msY-Net which processes two patches -- a central patch ($\times1$) and a context patch with the same patch dimension ($256 \times 256$) but larger FOV ($\times4$) -- in two parallel encoders.
To take the parallel encoder paths of the msY-Net and its increase in learnable parameters into consideration, we doubled the learning rate for all msY-Net experiments.
Since the msY-Net implementation is based on ResNet-18, we will compare it to ResNet-18 versions of the U-Net and DeepLabV3 architectures.

We note that \cite{schmitz2021} validate the performance on randomly sampled image crops and on full-scale WSIs ($\mathit{ds}=1$). For a competing benchmark, we therefore retrained the msY-Net maintaining our training procedure and evaluated all models on the identical WSI data.

\vspace{+3pt}
\noindent\textbf{Model training}
All models are based on ImageNet \cite{russakovsky2015imagenet} pre-trained ResNet \cite{he2015resnet} encoders. 
Our default MAF is configured with an embedding size of $D=1024$, the MHA with 8 heads and a hidden dimension of $d=128$, each, sinusoidal 1D position encodings and a neighbourhood radius $k=8$.

\vspace{+3pt}
\noindent\textit{Patch sampling:} 
To cope with the class imbalance and the excessive number of patches, we use a patch-sampling mechanism for the train and validation phase: To avoid losing important information of infrequent classes, we randomly draw at most 100 patches per WSI and tissue class -- assigning each patch its predominant tissue class. For comparison, we ensure that all models are trained with the same patch sample, for each fold respectively.
At test time, we evaluate the model on all patches of each WSI in the test set.

\vspace{+3pt}
\noindent\textit{Data augmentation:}
For all models, we restrict to colour jitter only since spatial augmentations (flipping, rotating, cropping) might lead to spatial inconsistencies between a central patch and its neighbourhood memory. For context patches in the msY-Net, we ensure the identical random augmentation as its central patch.

\vspace{+3pt}
\noindent\textit{Optimization:}
We use SGD with a momentum of 0.9 and a learning rate of 0.0001 (0.0002 for msY-Net) with exponential learning rate decay of $\beta=0.95$. Each model is trained with a batch size of 32 for $100$ epochs and its validation loss is determined after each epoch. Early stopping is applied if the validation loss does not further decrease for $10$ epochs. At the best validation loss, the model is evaluated.

\vspace{+3pt}
\noindent\textit{Neighbourhood memory:}
The memory $\mathcal{M}$ of the MAF is updated at the start of each epoch. All patches of all WSIs are compressed by the model in PyTorch evaluation mode with disabled gradient calculation (see \ref{subsec_mem}). We deviate from the patch sampling to ensure a complete neighbourhood for every sampled patch querying the memory.

\vspace{+3pt}
\noindent\textbf{Implementation}
All models were implemented in PyTorch 1.10.0. using \textit{Segmentation Models} \cite{yakubovskiy2019}. For the MAF, we implemented a combined memory $\mathcal{M}$ for all WSIs in a data set, for each phase respectively, using PyTorch Tensors to allow an efficient, simultaneous access of all neighbourhood embeddings for one batch (cross-WSI). The memory is deployed on the GPU memory (optionally on the CPU memory if size exceeds VRAM). Depending on the number of WSIs in the train set, the patch dimensions $n_x$ and $n_y$, the embedding size $D$ and the neighbourhood radius $k$, the physical memory size of $\mathcal{M}$ can be determined as:
\begin{equation*}
	|\mathcal{W}| \cdot (2k + n_x) \cdot (2k + n_y) \cdot D \cdot \SI{4}{\byte\per FP32}
\end{equation*}
Thus, the complete WSI train set can be compressed to a memory size feasible for a modern GPU. E.g. for the RCC data set at $\mathit{ds}=16$ and $k=8$, the physical size of the embedding memory $\mathcal{M}$ for the train set is 
\begin{equation*}
	112 \cdot (2 \cdot 8 + 65) \cdot (2 \cdot 8 + 36) \cdot 1024 \cdot \SI{4}{\byte\per FP32} = \SI{1.80}{\giga\byte}
\end{equation*}
For the msY-Net, we used the model implementation of the official repository.
All experiments were run on a \SI{48}{\giga\byte} NVIDIA RTX A6000.

\vspace{+6pt}
\noindent\textbf{Validation}
We run each experiment with 5-fold cross validation strategy (CV) -- splitting on the level of WSI -- resulting in 140/35 WSIs in the training/test set per fold for the RCC data set and 16/4 WSIs in the training/test set per fold for the CY16 data set. For early stopping, we use $20 \%$ of the training set -- resulting in 28 and 4 WSIs in the validation set, respectively. For comparison between all experiments, we ensure the same splits between all models, respectively per fold. 

\vspace{+6pt}
\noindent\textbf{Evaluation metrics}
We measure the semantic segmentation performance with the micro-average Dice Similarity Coefficient ($\mathit{DSC}$) for each tissue class $c$ as $\mathit{DSC}_c = \overline{m}_{\mathrm{NaN}}^{\mathrm{W}}(\mathit{DSC}(c,w))$ and in total as $\mathit{DSC}_{\mathrm{total}} = \overline{m}_{\mathrm{NaN}}^{\mathcal{W}}( \overline{m}_{\mathrm{NaN}}^{\mathrm{C}}(\mathit{DSC}(c,w)))$ with $\overline{m}_{\mathrm{NaN}}$ as the mean function for all defined $\mathit{DSC}$s. 

We estimate the model performance with the mean $\mathit{DSC}$ over all folds defined as $\overline{\mathit{DSC}}$ alongside with its standard deviation.

\vspace{+10pt}
\section{Results}
\vspace{+10pt}
\subsection{Patch-based Baselines}
To benchmark the performance gain of enabling the MAF, we first determined the performance of patch-based segmentation models and studied the effect of expanding the FOV (increasing $\mathit{ds}$) -- without enabling the MAF. For the RCC data set, both, U-Net and DeepLabV3, in combination with both encoders, ResNet-18 and ResNet-50, consistently performed best at $\mathit{ds}=16$ (see Supplementary Table \ref{rcc_scale}). The combination of DeepLabV3 and ResNet-50 yielded the best results with $0.50~\overline{\mathit{DSC}}_{total}$.
Zooming out too far (less details -- larger FOV) deteriorates the performance significantly.
For the CY16 data set, we observe a strong performance ($0.73~\overline{\mathit{DSC}}_{Tumour}$) of patch-based models at $\mathit{ds}=2$ using the U-Net with ResNet-18 (see Supplementary Table \ref{cy16_scale}).
We then freezed $\mathit{ds}=16$ for all RCC MAF experiments and $\mathit{ds}=2$ for all CY16 MAF experiments.

\subsection{MAF model variations}

Next, we studied the effect of altering architectural concepts in our MAF -- based on the RCC data set and the best baseline setup with DeepLabV3 and ResNet-50 encoder at $\mathit{ds}=16$. We analyze the effect of the memory embedding size, position encodings, helper loss and the neighbourhood size in a gradual manner. We start with a default MAF experiment setup using $D=1024$, sinusoidal position encodings, $\lambda=0$ (no helper loss) and $k=8$, and gradually alter the concepts. %Table \ref{mod_var} shows the results.

\vspace{+3pt}
\noindent\textbf{Effect of memory embedding size}
We analysed the impact of the embedding dimension $D$ and hypothesize that a higher $D$ allows compressing more information into the memory. Table \ref{mod_var} shows that the performance is best for $D=1024$. Increasing it to $2048$ deteriorates the performance -- assuming that too large embeddings might lead to overfitting. On the other hand, decreasing $D$ to 512 also deteriorate the performance - assuming to little semantic space for storing the compression. For all following experiments, we therefore use $D=1024$.

\vspace{+6pt}
\noindent\textbf{Effect of positional encoding}
Subsequently, we analysed the effect of positional encodings. Without, the attention mechanism of the central patch is not aware of the spatial relationships with its neighbours.
We compare the performance of applying 1D sinusoidal embeddings \cite{vaswani2017attention} with using no position embeddings and applying relative 2D learnable embeddings \cite{ramachandran2019stand}.
Table \ref{mod_var} shows that the relative 2D learnable embeddings add most benefit, followed by 1D sinusoidal embeddings. Using no position embeddings performs worst -- still outperforming the baseline though.

\vspace{+6pt}
\noindent\textbf{Effect of helper loss}
We hypothesized that a loss $\mathcal{L}_{cls}$ predicting the central patch's class distribution from the context embedding $a$ might guide the MHA module. Therefore, we add $\lambda \cdot \mathcal{L}_{cls}$ to $ (1-\lambda) \cdot \mathcal{L}_{seg}$. Table \ref{mod_var} shows the results for different $\lambda$. We observe the best performance for $\lambda=0.2$ and freeze it for following experiments (referred to $\text{MAF}^+$).

\begin{table*}[ht]
	\captionsetup{justification=raggedright, singlelinecheck=off}
	\centering
	\begin{threeparttable}
		\caption{Analysis of variations for DeepLabV3+MAF with $\mathit{ds}=16$ and $k=8$ on RCC.}
		
		\begin{tabular}{@{}cccccccccccc@{}}\toprule
			\multicolumn{3}{c}{$D$} & \phantom{a} & \multicolumn{3}{c}{Pos. enc.} & \phantom{a} & \multicolumn{3}{c}{$\mathcal{L}_{cls} \text{ with } \lambda$} & $\overline{\mathit{DSC}}_{total}$\\
			\cmidrule{1-3}  \cmidrule{5-7} \cmidrule{9-11}
			512 & 1024 & 2048 && no pos. & sin. 1D pos. & rel. 2D pos. && 0.2 & 0.5 & 0.8\\ \midrule
			\checkmark & -                                  & -          &   & -          & \checkmark & -                                  &   & -          & -          & -                                  & 0.5315          \\
			-          & \textcolor{OliveGreen}{\checkmark} & -          &   & -          & \checkmark & -                                  &   & -          & -          & -                                  & \textbf{0.5470} \\
			-          & -                                  & \checkmark &   & -          & \checkmark & -                                  &   & -          & -          & -                                  & 0.5398          \\
			\midrule
			-          & {\checkmark}                       & -          &   & \checkmark & -          & -                                  &   & -          & -          & -                                  & 0.5312          \\
			-          & {\checkmark}                       & -          &   & -          & -          & \textcolor{OliveGreen}{\checkmark} &   & -          & -          & -                                  & \textbf{0.5517} \\
			\midrule
			-          & {\checkmark}                       & -          &   & -          & \checkmark & -                                  &   & \textcolor{OliveGreen}{\checkmark} & -          & -                                  & \textbf{0.5672}          \\
			-          & {\checkmark}                       & -          &   & -          & \checkmark & -                                  &   & -          & \checkmark & -                                  & 0.5662          \\
			-          & {\checkmark}                       & -          &   & -          & \checkmark & -                                  &   & -          & -          & {\checkmark} & 0.5589 \\
			\bottomrule
		\end{tabular}
		\label{mod_var}
	\end{threeparttable}
\end{table*}

\begin{figure}[!t]
	\centering
	\includegraphics[width=0.4\textwidth]{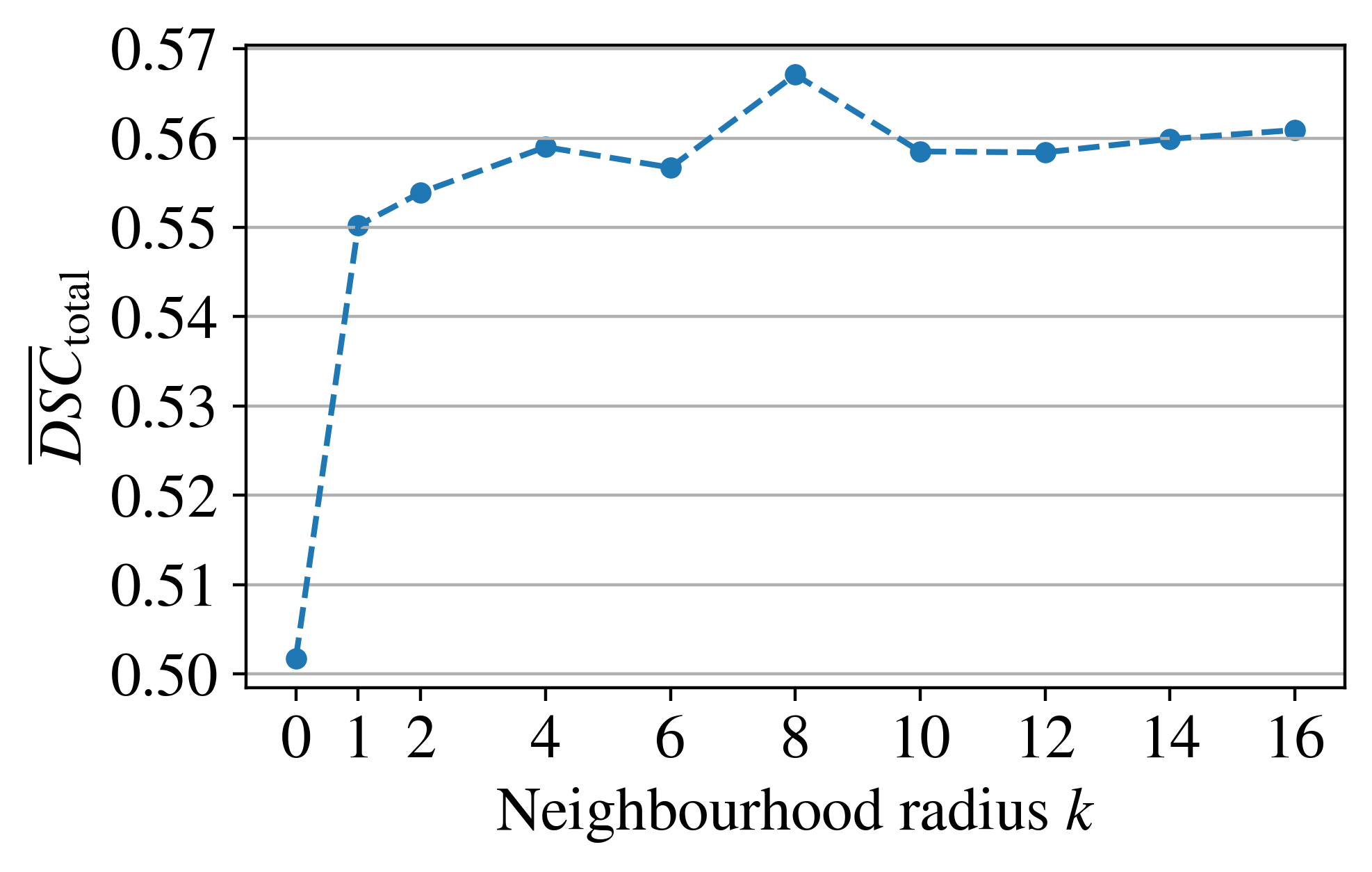}
	\caption{$\overline{\mathit{DSC}}_{\mathrm{total}}$ at different neighbourhood radii for RCC with DeepLabV3+$\text{MAF}^+$ at $\mathit{ds}=16$.}
	\label{k_plot}
\end{figure}

\vspace{+6pt}
\noindent\textbf{Effect of neighbourhood radius}
To better understand the effect of the neighbourhood attention on the segmentation performance, we compared the DeepLabV3+$\text{MAF}^+$ performance using different neighbourhood radii $k$ (see Fig. \ref{k_plot}). Note that $k=0$ equals DeepLabV3 w/o MAF.
While most performance gain is observed from $k=0 \rightarrow k=1$ (using no neighbourhood $\rightarrow$ adjacent patches only), we observe a further increase of performance until $k=8$ followed by a subtle lower performance with $k>8$. Still, all versions of $\text{MAF}^+$ with $k>0$ outperform the baseline significantly.
We decided to freeze $k=8$ for all following experiments.
The resulting neighbourhood size is comparable to a FOV expansion from $\mathit{ds}=16$ to $\mathit{ds}=512$, thus, increasing the FOV by $\times32$.

\begin{table*}\centering
	%\ra{1.3}
	\caption{RCC -- 5-fold CV $\overline{\mathit{DSC}}_c$ w/o and w/ $\text{MAF}^{+}$ with $\mathit{ds}=16$ and $k=8$ (\textit{Angioinv.} excluded since all architectures failed to detect). *Non-Tumour ** $\overline{\mathit{DSC}}_{\mathrm{total}}$}
	\begin{tabular*}{\textwidth}{@{}c@{\extracolsep{\fill}}lgc@{\hskip -0.025in}cc@{\hskip 0.1in}gc@{\hskip -0.025in}c@{}}\toprule
		& \thead{Tissue \\ Class} & U-Net & \thead{U-Net \\ + $\text{MAF}^{+}$} & $\Delta$ && DeepLabV3 & \thead{DeepLabV3 \\ + $\text{MAF}^{+}$} & $\Delta$ \\ \midrule
		\multirow{5}{*}{\rotatebox[origin=c]{90}{Tumour}} & Vital & $0.90 \pm 0.02$ & $0.91 \pm 0.02$ & \textcolor{OliveGreen}{+0.01} && $0.90 \pm 0.02$ & $0.91 \pm 0.02$ & \textcolor{OliveGreen}{+0.01} \\
		& Regression & $0.46 \pm 0.04$ & $0.49 \pm 0.04$& \textcolor{OliveGreen}{+0.03} && $0.47 \pm 0.04$ & $0.51 \pm 0.04 $& \textcolor{OliveGreen}{+0.04}\\
		& Necrosis & $0.23 \pm 0.06$ & $0.22 \pm 0.06$ & $-0.01$ && $0.24 \pm 0.07$ & $0.33 \pm 0.11 $& \textcolor{OliveGreen}{+0.09} \\
		& Bleeding & $0.22 \pm 0.06$ & $0.18 \pm 0.02$ & $-0.04$ && $0.25 \pm 0.04$ & $0.29 \pm 0.05 $ & \textcolor{OliveGreen}{+0.04} \\
		%& Angioinv. & $0.0 \pm 0.0$ & $0.0 \pm 0.0$ & +0.00 && $0.0 \pm 0.0$ & $0.0 \pm %0.0$ & %+0.00\\
		& Capsule & $0.25 \pm 0.02$ & $0.32 \pm 0.02$ & \textcolor{OliveGreen}{+0.07} && $0.29 \pm 0.23$ & $0.36 \pm 0.01$ & \textcolor{OliveGreen}{+0.07} \\
		& Cyst & $0.00 \pm 0.00$ & $0.00 \pm 0.00$ & +0.00 && $0.02 \pm 0.03$ & $0.05 \pm 0.11$ & \textcolor{OliveGreen}{+0.03} \\  
		\hline
		\multirow{3}{*}{\rotatebox[origin=c]{90}{Non-T*}} & Cortex & $0.51 \pm 0.07$ & $0.54 \pm 0.08$ & \textcolor{OliveGreen}{+0.03} && $0.52 \pm 0.07$ & $0.57 \pm 0.38$ & \textcolor{OliveGreen}{+0.05} \\
		& Mark & $0.28 \pm 0.05$ & $0.30 \pm 0.04$ & \textcolor{OliveGreen}{+0.02} && $0.30 \pm 0.05$ & $0.41 \pm 0.06$ & \textcolor{OliveGreen}{+0.11} \\
		& Extrarenal & $0.66 \pm 0.02$ & $0.68 \pm 0.02$ & \textcolor{OliveGreen}{+0.02} && $0.67 \pm 0.03$ & $0.69 \pm 0.02$ & \textcolor{OliveGreen}{+0.02} \\
		\midrule
		& \textbf{WSI**} & $0.48 \pm 0.01$ & $0.50 \pm 0.02$ & \textcolor{OliveGreen}{+0.02} && $0.50 \pm 0.01$ & $ \textbf{0.57} \pm 0.02$ & \textcolor{OliveGreen}{+0.07} \\
		\bottomrule
	\end{tabular*}
	\label{rcc_subclass_performance}
	%\vspace{-4mm}
\end{table*}

\begin{table*}\centering
	%\ra{1.3}
	\caption{Context-integrating contribution of $\text{MAF}^+$ compared with msY-Net (all based on ResNet-18) *$\overline{\mathit{DSC}}_{\mathrm{total}}$ **$\overline{\mathit{DSC}}_{\mathrm{Tumour}}$}
	\begin{tabular*}{\textwidth}{@{}lc@{\extracolsep{\fill}}gc@{\hskip -0.025in}cc@{\hskip -0.025in}cc@{\hskip -0.025in}c@{}}\toprule
		\thead{Data Set} & $\mathit{ds}$ & U-Net & msY-Net & $\Delta$ & \thead{U-Net \\ + $\text{MAF}^+$} & $\Delta$ &  \thead{DeepLabV3 \\ + $\text{MAF}^+$} & $\Delta$   \\ \midrule
		RCC* & 16 & $0.45 \pm 0.02$ & $0.46 \pm 0.02$ & $\textcolor{OliveGreen}{+0.01}$ & $0.47 \pm 0.02$ & $\textcolor{OliveGreen}{+0.02}$ & $\textbf{0.54} \pm 0.01$ & \textcolor{OliveGreen}{+0.09} \\
		CY16** & 2 & $0.73 \pm 0.06$ & $0.79 \pm 0.08$ & $\textcolor{OliveGreen}{+0.06}$ & $0.76 \pm 0.09$ & $\textcolor{OliveGreen}{+0.03}$ & $\textbf{0.80} \pm 0.09$ & $\textcolor{OliveGreen}{+0.07}$ \\
		\bottomrule
	\end{tabular*}
	\label{cam_performance}
	%\vspace{-8mm}
\end{table*}

\subsection{RCC baseline}
Using the best performing $\text{MAF}^+$ setting (patch embedding dimension $D=1024$, sin. 1D position encoding, helper loss with $\lambda=0.2$, neighbourhood radius $k=8$), we compare its class-wise performance on the RCC data set against the patch-based baselines in Table \ref{rcc_subclass_performance}, both for U-Net and DeepLabV3 base architectures. We found both -- U-Net+$\text{MAF}^+$ ($0.50~\overline{\mathit{DSC}}_\mathrm{total}$) and DeeplabV3+$\text{MAF}^+$ ($0.57~\overline{\mathit{DSC}}_\mathrm{total}$) -- outperforming their baseline by $0.02$ for U-Net and $0.07$ ($+14\%$) for DeepLabV3. The former (U-Net+$\text{MAF}^+$) improves best in the segmentation of \textit{Capsule} tissue by $0.07$ but disimproves in the segmentation of \textit{Tumour Bleeding} by $0.04$.
The latter (DeeplabV3+$\text{MAF}^+$) shows improvements for every subtype of tissue with the largest performance gain for \textit{Mark} tissue by $0.11$. All methods -- baseline and $\text{MAF}^+$ -- could not detect the minority classes \textit{Angioinvasion} yielding $0.00~\overline{\mathit{DSC}}_\mathrm{Angioinvasion}$.

\subsection{Context benchmark}
Next, we benchmark the context integration effect of the MAF with the effect of the msY-Net (based on ResNet-18) proposed by \cite{schmitz2021} on the internal RCC data set and on the public CY16 data set. For comparison, we changed all encoders to ResNet-18 (Table \ref{cam_performance}).

For \textbf{RCC} with $\mathit{ds}=16$, the U-Net baseline reaches $0.45~\overline{\mathit{DSC}}_\mathrm{total}$. The msY-Net model marginally outperforms the baseline with $0.46~\overline{\mathit{DSC}}_\mathrm{total}$ ($+0.01$). The U-Net+MAF$^+$ scores at $0.47~\overline{\mathit{DSC}}_\mathrm{total}$ ($+0.02$). DeepLabV3+MAF$^+$ ($0.54~\overline{\mathit{DSC}}_\mathrm{total}$) significantly outperforms the baseline by 0.09 (+\%20) and the msY-Net by 0.08 (+17\%).

For \textbf{CY16} with $\mathit{ds}=2$, we report the \textit{Tumour} $\overline{\mathit{DSC}}$ following \cite{schmitz2021}. The U-Net baseline scores at $0.73~\overline{\mathit{DSC}}_{\mathrm{Tumour}}$. Our msY-Net model outperforms the baseline by 0.06 with $0.79~\overline{\mathit{DSC}}_{\mathrm{Tumour}}$. We observe the U-Net+MAF$^+$ outperforming the U-Net baseline by 0.03 with $0.76~\overline{\mathit{DSC}}_{\mathrm{Tumour}}$, however does not reach the msY-Net performance.
Finally, we show that DeepLabV3+MAF$^+$ scores best with $0.80~\overline{\mathit{DSC}}_{\mathrm{Tumour}}$ marginally outperforming the msY-Net.

\subsection{Qualitative results}

\begin{figure*}[!tp]
	\centering
	\includegraphics[width=\textwidth]{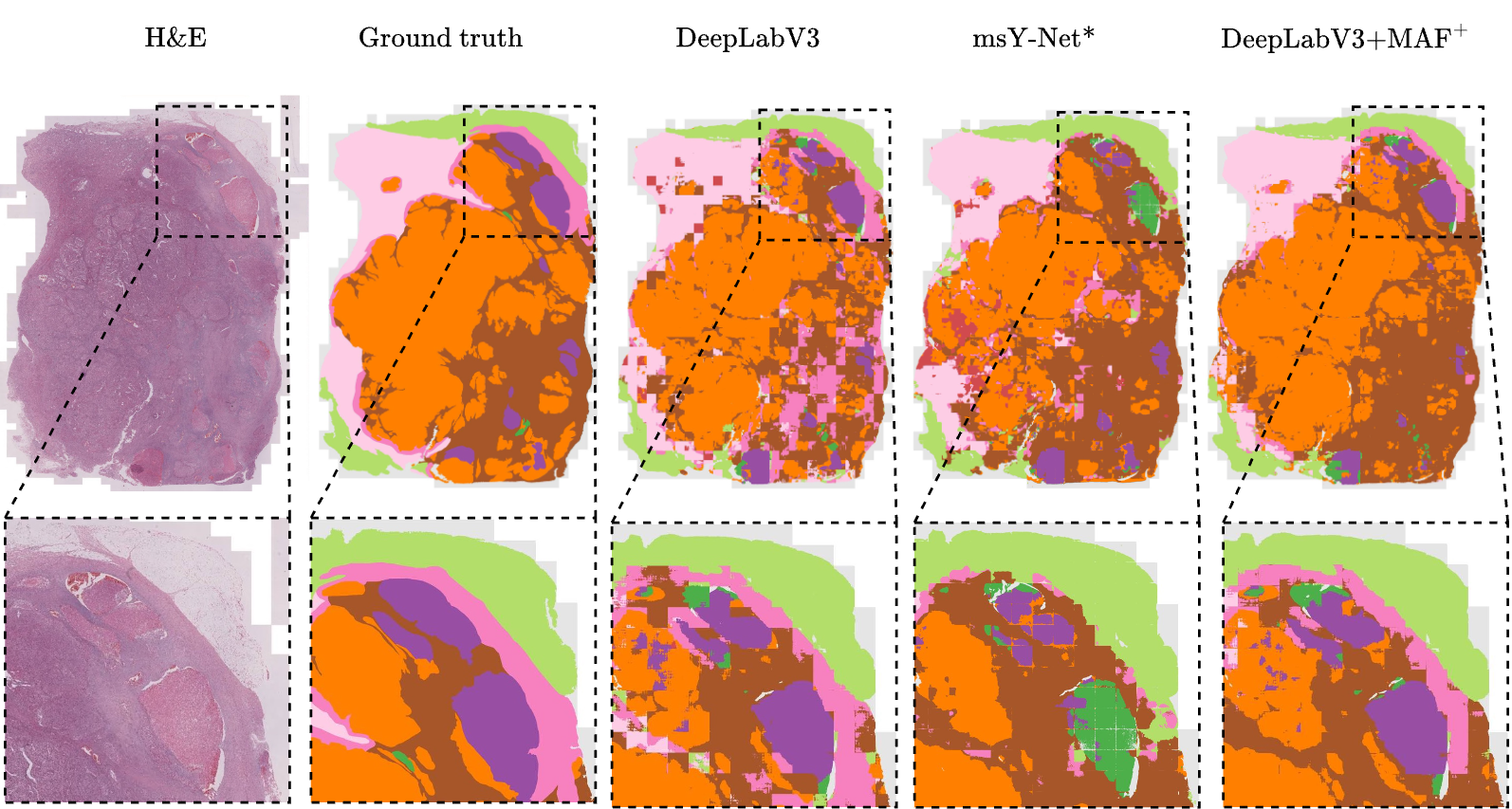}
	\caption{Comparison of RCC segmentation with $\mathit{ds}=16$ at different magnifications. *based on ResNet-18} 
	\label{wsi_quali}
\end{figure*}

\begin{figure*}[!tp]
	\centering
	\includegraphics[width=\textwidth]{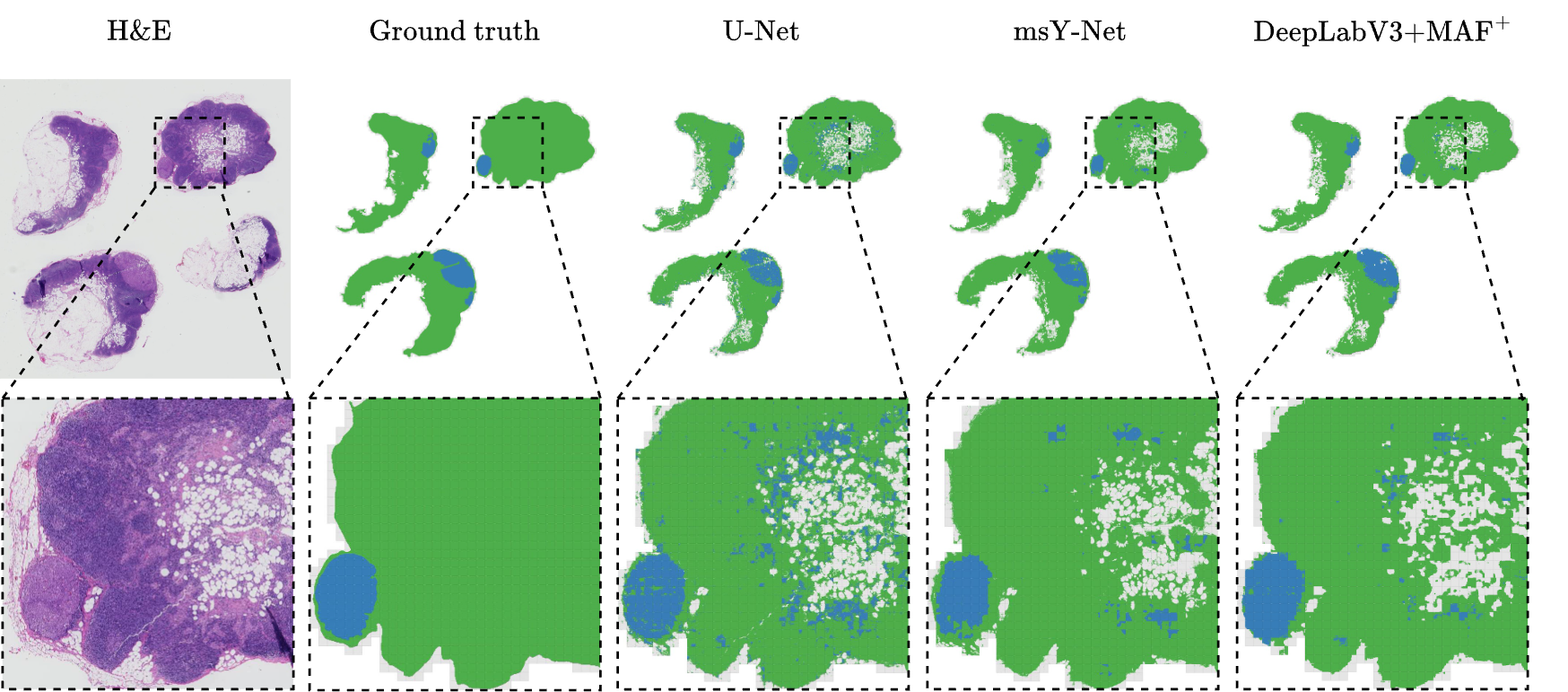}

	\caption{Comparison of CY16 segmentation with $\mathit{ds}=2$ at different magnifications (all based on ResNet-18).} 
	\label{cy_wsi_quali}
\end{figure*}
\begin{figure*}[!t]
	\centering
	\includegraphics[width=0.8\textwidth]{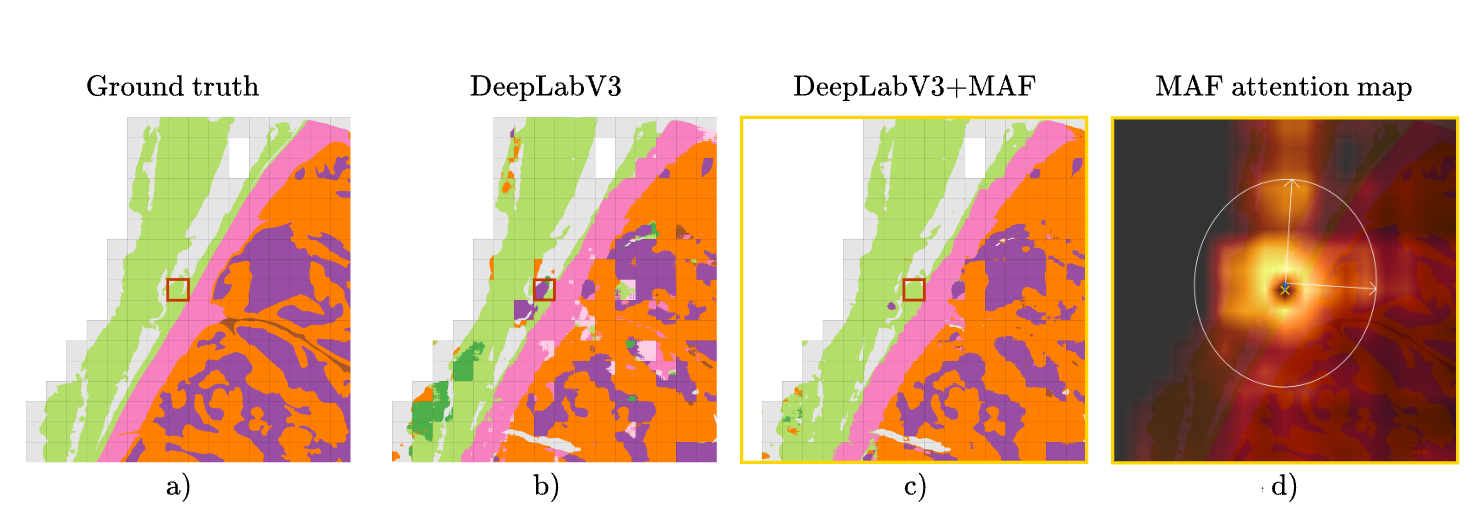}
	\caption{Example of RCC segmentation details and attention map of the central patch. a) Ground truth d) Segmentation prediction of DeepLabV3 c) Segmentation prediction of DeepLabV3+MAF b) Attention map with respect to the central patch. \textcolor{redd}{$\square$} Central patch \textcolor{goldyellow}{$\square$} Attended neighbourhood of central patch}
	\label{patch_attention}
\end{figure*}
\begin{figure*}[!htb]
	\centering
	\includegraphics[width=\textwidth]{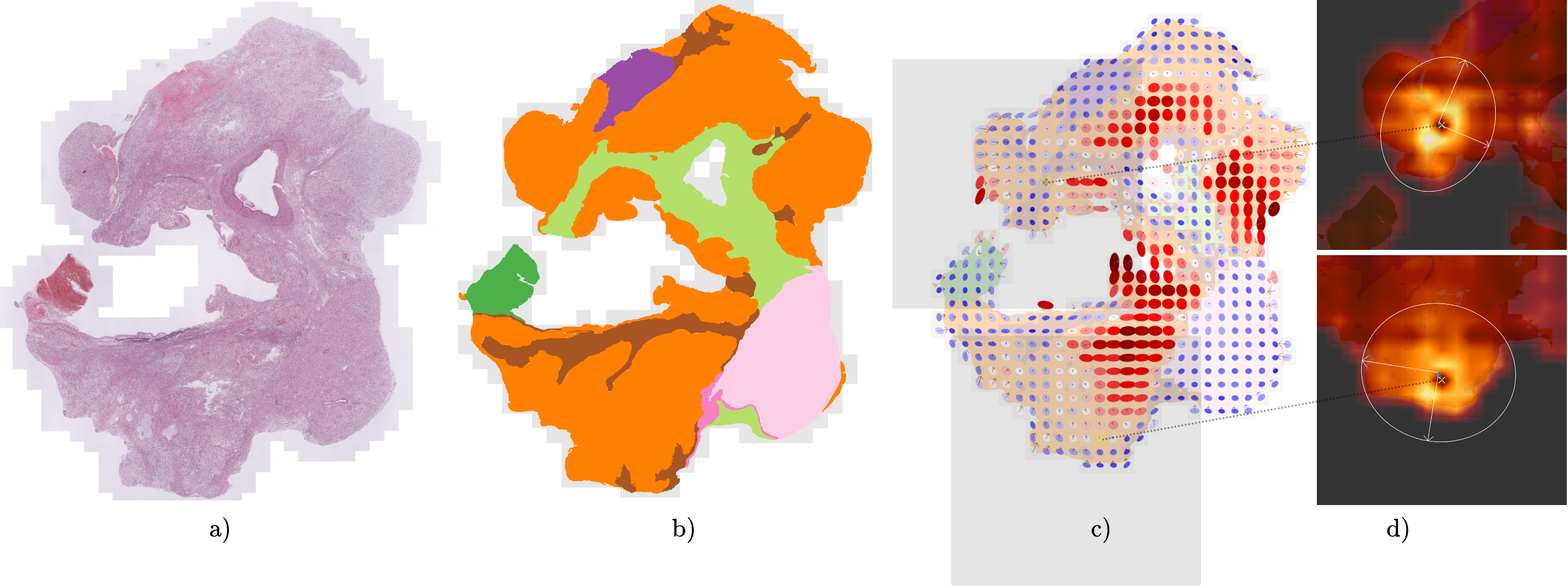}
	\caption{Example of aggregated attention view: a) Tissue b) Annotations c) Segmentation prediction with aggregated attention view. For each patch $p^c$, we determine its characteristic ellipse from the attention of the neighbourhood. We colourize each ellipse with respect to its deviation of area to the mean area (\textcolor{red}{$\blacksquare$}$\text{: } >\text{mean area} \rightarrow \text{long reach attention}$, \textcolor{blue}{$\blacksquare$}$\text{: }< \text{mean area} \rightarrow \text{close reach attention}$) and plot its miniature version for each patch. d) Attention heatmaps and its characteristic ellipse for a central patch $p^c$ (\textcolor{goldyellow}{$\times$}).} 
	\label{attention_analysis}
\end{figure*}

To visually perceive the benefit of the MAF, we show an example of an RCC segmentation of one WSI based on DeepLabV3+$\text{MAF}^+$ with $k=8$ and $\mathit{ds}=16$ in Fig. \ref{wsi_quali} and compare it to a patch-based DeepLabV3 with $\mathit{ds}=16$ and the msY-Net with $\mathit{ds}=16$ for the centre patch and $\mathit{ds}=64$ for the context patch.
Zooming in reveals that some patch segmentations of DeepLabV3 without the MAF are entirely wrong -- resulting in a scattered, non-cohesive segmentation map. The msY-Net improves in detecting more cohesive tissue but is more error-prone for tissue confusion. Integrating the MAF almost solves this issue. The attention mechanism identifies cohesive tissue much better. We provide more qualitative results for the RCC data set in the supplement (see Supplementary Fig. \ref{add_wsi_quali}). Fig. \ref{cy_wsi_quali} shows a segmentation result for the CY16 data set. While the U-Net patch-based approach wrongly detects scattered tumourous tissue patches, both -- msY-Net and DeeplabV3+MAF$^+$ -- improve their precision and are more aware of coherent tissue sections.

\subsection{Analysis of attention}

Enabling the attention mechanism shows an improvement in segmentation performance since context information can be retrieved from the neighbourhood. To better understand which information is used, we visualize the attention scores over the queried neighbourhood. All attention visualizations are based on DeepLabV3+MAF with learned 2D position encodings. 

\vspace{+3pt}
\noindent\textbf{Attention maps}
We enlarge an RCC segmentation result and visualize a cut-out in size of the corresponding neighbourhood -- here $k=8$. For each central patch, we can retrieve the resulting attention values of the MHA in the MAF. We average the values over all 8 heads and highlight the neighbourhood with respect to their relative attention importance. Fig. \ref{patch_attention} shows the resulting attention map. Interestingly, we can observe close-by neighbours attracting more attention. Also, the attention is not concentric to the central patch but rather biased by the patches' tissue classes. The segmentation of this patch without the MAF is of lower quality.

\vspace{+3pt}
\noindent\textbf{Attention views}
To further understand the attention mechanism, we create a view of aggregated attention maps over a WSI (see. Fig. \ref{attention_analysis}). We first average the attention scores over all heads and yield an aggregated 2D attention map with scores summing up to 1. Subsequently, for each central patch $p^c$, we fit the parameters of a bivariate normal distribution using least-squares on the attention scores -- for illustration purposes, simplified assuming the attention scores to form an empirical bivariate normal distribution around the central patch.
Fig. \ref{attention_analysis} d) shows two example central patches and their attending neighbourhood with its characteristic $90\%$ confidence ellipse (axis in the direction of Eigenvectors scaled by the square root of the Eigenvalue times $\chi_{\alpha=0.9}^2({r=2})$, respectively). We also show the shift of the ellipse centre from the patch centre as a blue arrow -- indicating the attention focus.
To come up with an aggregated attention view, for each central patch, we plot a miniature version (Fig. \ref{attention_analysis} c) ) of its characteristic ellipse -- colouring the ellipse area by its deviation from the mean area (\textcolor{red}{$\blacksquare$}$\text{: } >\text{mean area} \rightarrow \text{long reach attention}$, \textcolor{blue}{$\blacksquare$}$\text{: }< \text{mean area} \rightarrow \text{close reach attention}$) and indicating its attention focus with a black arrow representing the ellipse centre shift. We provide more attention views in the supplement (see Supplementary Fig. \ref{add_att}).

\section{Discussion and conclusions}

We proposed a semantic segmentation extension for WSIs using a patch neighbour attention mechanism that queries the neighbouring tissue context from an embedding memory bank and infuses context embeddings into bottleneck feature maps. We showed that our approach is superior to patch-based segmentation algorithms and even outperforms SOTA context-integrating algorithms in a multi-class cancer data set. Our MAF facilitates a much wider FOV to access context information compared to SOTA -- while simultaneously processing details on a patch level. In addition, the MAF is able to learn, first, a compressed form of the context and, second, to selectively attend relevant context information. We show that the MAF is robust in terms of neighbourhood size.

One could observe that the performance boost of the MAF on the RCC segmentation is superior to the CY16 segmentation. This is reasonable as first, the RCC segmentation task is more complex due to its multiple subtypes of tissue and second, the tumour identification of macrometastasis (CY16) can be achieved at cell level from a pathologists perspective, thus less favourable for the MAF. We conclude that the MAF is more beneficial to the segmentation of complex tissue structures where a human pathologist needs to make use of context information (zooming out and considering surrounding tissue context). Our visualization of the attention maps -- mimicking a pathologist's view -- supports this conclusion.

From a pathologist's view, the attention views show interesting characteristics: The attention reach of \textit{Vital Tumour} exceeds all other tissue types. At the tumour border lamella, we can see an intensified reach of attention clearly with a focus on the border -- indicating the detection of the tumour border.
Also, we can observe an increase of attention reach for areas with subtle type of tissue borders (e.g. \textit{Cortex} $\rightarrow$ \textit{Mark}) -- indicating the intensified usage of attention for regions with clear context needs. 
We can also observe intensified attention reach for tissue with morphological heterogeneity -- e.g. in \textit{Extrarenal} tissue -- indicating the access to more context information in case of indecisiveness using the centre patch information only. On the other hand, the attention reach is limited for tissue regions with morphological homogeneity e.g. \textit{Extrarenal} fat tissue. As stated by our pathologist, we can observe similar usage of context information compared to them.

Applying the MAF to DeepLabV3 shows a larger benefit than to U-Net and we assume that architectural elements (e.g. atrous spatial pyramid pooling) favour the context fusion. Different fusing mechanisms and encoder-decoder architectures should therefore be studied. In this work, we applied one MHA layer only. We plan to change it to a more receptible Transformer encoder to extend the attention capability.
In future work, we believe the memory can be exploited even further to better process context information -- e.g. by integrating the context embeddings into different decoder levels or fully Transformer-based architectures. Also, we believe that it will help with memory-expensive 3D segmentation tasks by applying the MAF on image slices. To tackle the trade-off between FOV and physical resolution, research about hierarchical memory attention mechanisms storing patches each at multiple FOVs seems a promising direction.

\section*{Acknowledgments}
This work received funding from \lq KITE' (Plattform für KI-Translation Essen) from the REACT-EU initiative (\url{https://kite.ikim.nrw/}) and HIDSS4HEALTH.
Jessica Schmitz and Jan Hinrich Bräsen were supported by German Ministry for Education and Research (BMBF 13GW0399B) and Jessica Schmitz, Jan Hinrich Bräsen, Nikolaos Vasileiadis, Philipp Ivanyi and Viktor Grünwald by the Wilhelm Sander Foundation.
We would like to thank E. Christians, M. Taleb-Naghsh and J. Jost for their excellent technical assistance.

%%%%%%%%%%%%%%   Bibliography   %%%%%%%%%%%%%%
\normalsize
\bibliographystyle{unsrtnat}
\bibliography{bibliography}

%%%%%%%%%%%%  Supplementary Figures  %%%%%%%%%%%%
\clearpage
\footnotesize
\section*{Supplementary material}

\begin{table*}
	\begin{minipage}{.5\linewidth}
		\caption{RCC baseline 5-fold CV $\overline{\mathit{DSC}}_{total}$ for different $\mathit{ds}$ (best $\mathit{DSC}$ in bold).}
		\begin{tabular}{@{}lccccc@{}}\toprule
			& \multicolumn{2}{c}{U-Net} & \phantom{a} & \multicolumn{2}{c}{DeepLabV3}\\
			\cmidrule{2-3} \cmidrule{5-6}
			$\mathit{ds}$       & ResNet-18     & ResNet-50     &   & ResNet-18     & ResNet-50     \\ \midrule
			$8$         & 0.44          & 0.45          &   & 0.47          & 0.47          \\
			\textbf{16} & \textbf{0.45} & \textbf{0.48} &   & \textbf{0.49} & \textbf{0.50} \\
			$32$        & 0.42          & 0.45          &   & 0.46          & 0.48          \\
			$64$        & 0.33          & 0.39          &   & 0.42          & 0.44          \\
			\bottomrule
		\end{tabular}
		\label{rcc_scale}
	\end{minipage}
	\begin{minipage}{.5\linewidth}
		\caption{CY16 baseline 5-fold CV $\overline{\mathit{DSC}}_{tumour}$ for different $\mathit{ds}$ (best $\mathit{DSC}$ in bold).}
		\begin{tabular}{@{}lccccc@{}}\toprule
			& \multicolumn{2}{c}{U-Net} & \phantom{a} & \multicolumn{2}{c}{DeepLabV3}\\
			\cmidrule{2-3} \cmidrule{5-6}
			$\mathit{ds}$       & ResNet-18     & ResNet-50     &   & ResNet-18     & ResNet-50     \\ \midrule
			1          & 0.68          & 0.68          &   & 0.74          & 0.73          \\
			\textbf{2}         & \textbf{0.73} & 0.72          &   & 0.76          & 0.75          \\
			4          & 0.71          & 0.73          &   & 0.76          & 0.76          \\
			8 & 0.72          & \textbf{0.75} &   & \textbf{0.77} & \textbf{0.78} \\
			16         & 0.63          & 0.72          &   & 0.73          & 0.76          \\
			\bottomrule
		\end{tabular}
		\label{cy16_scale}
	\end{minipage}
\end{table*}

\begin{figure*}
	\centering
	\includegraphics[width=\textwidth]{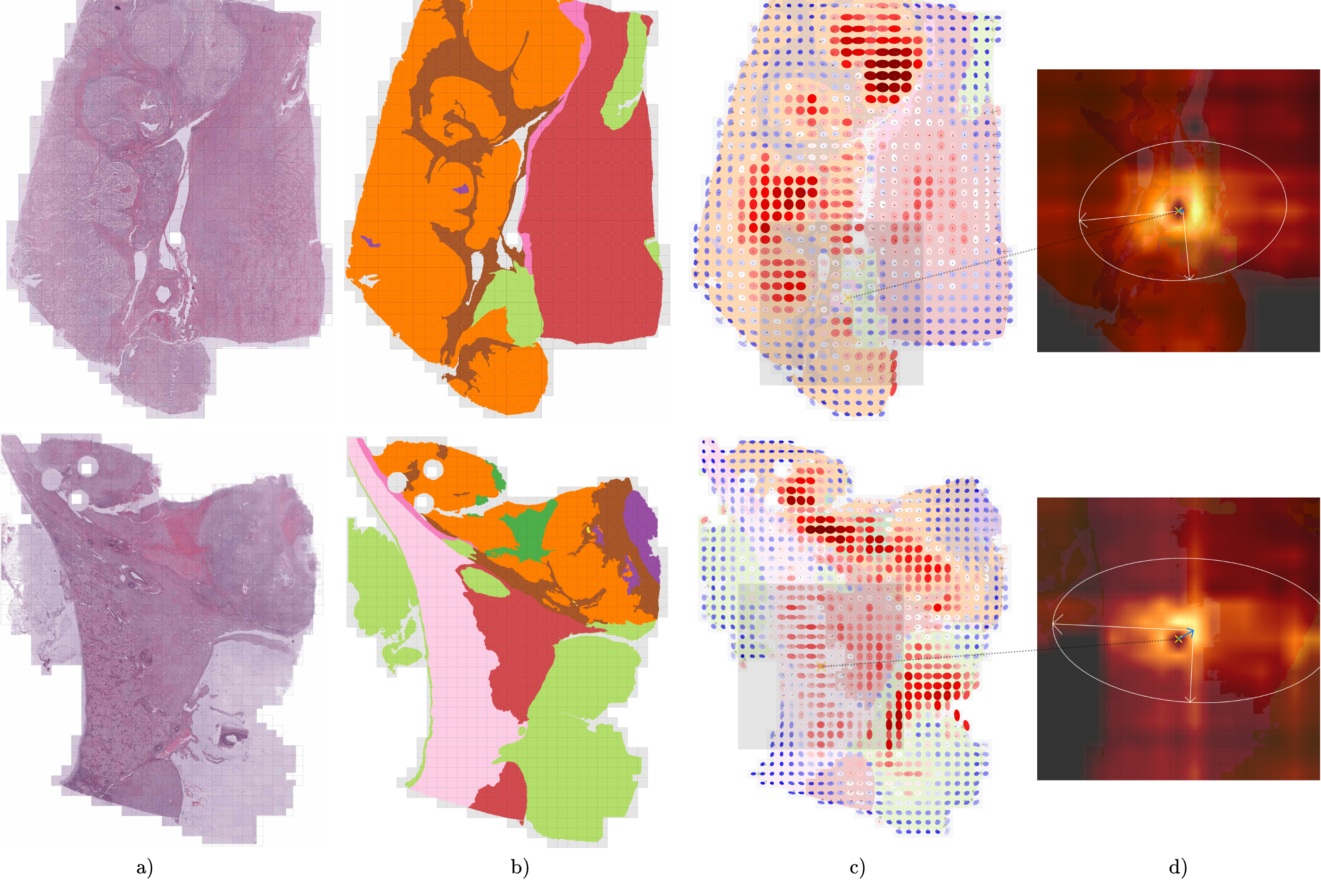}
	\caption{Additional attention views of RCC segmentations: a) Tissue b) Annotations c) Segmentation prediction with aggregated attention view d) Heatmap of neighbourhood attention for a central patch (\textcolor{goldyellow}{$\times$}).} 
	\label{add_att}
\end{figure*}

\begin{figure*}
	\centering
	\includegraphics[width=\textwidth]{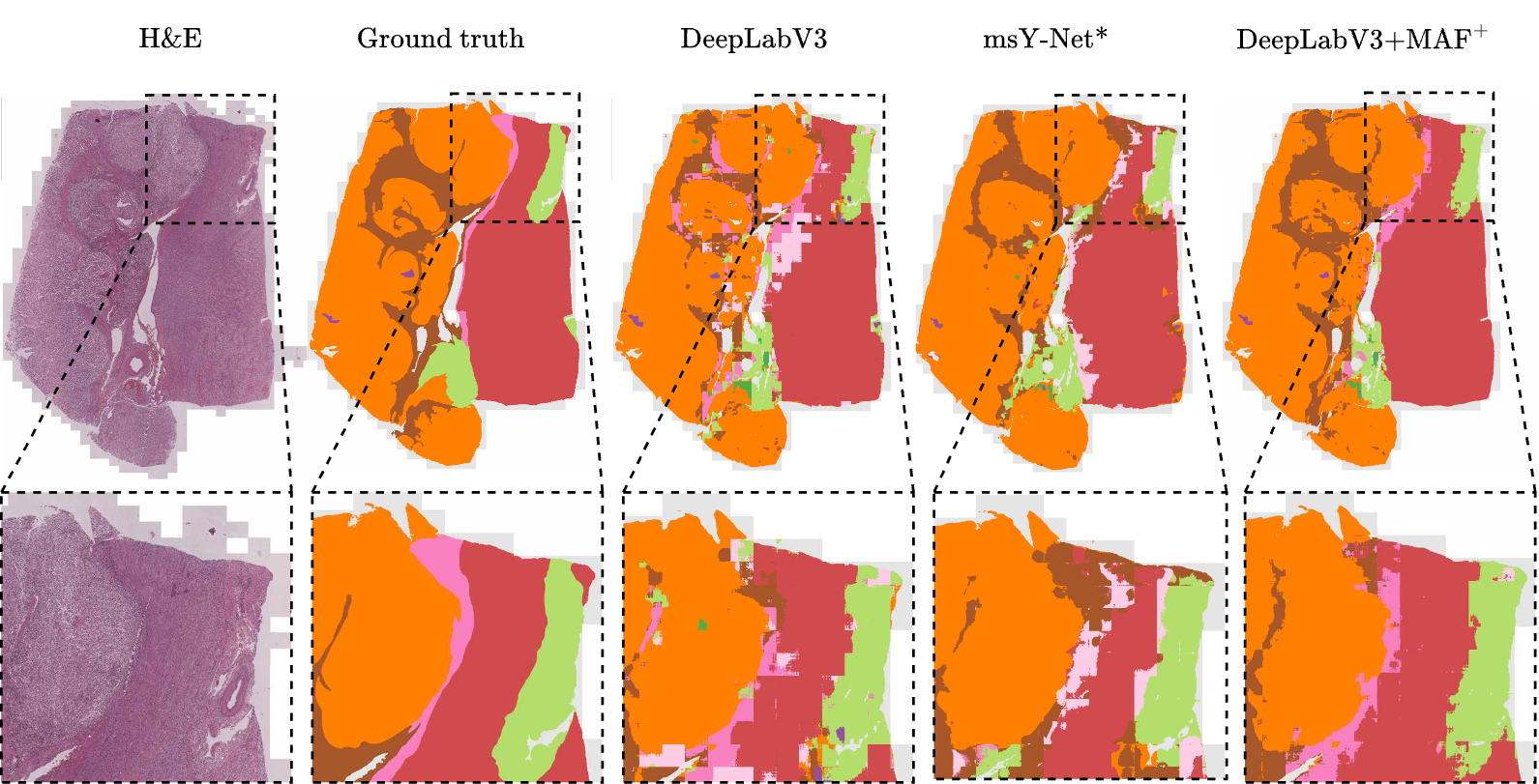}
	\includegraphics[width=\textwidth]{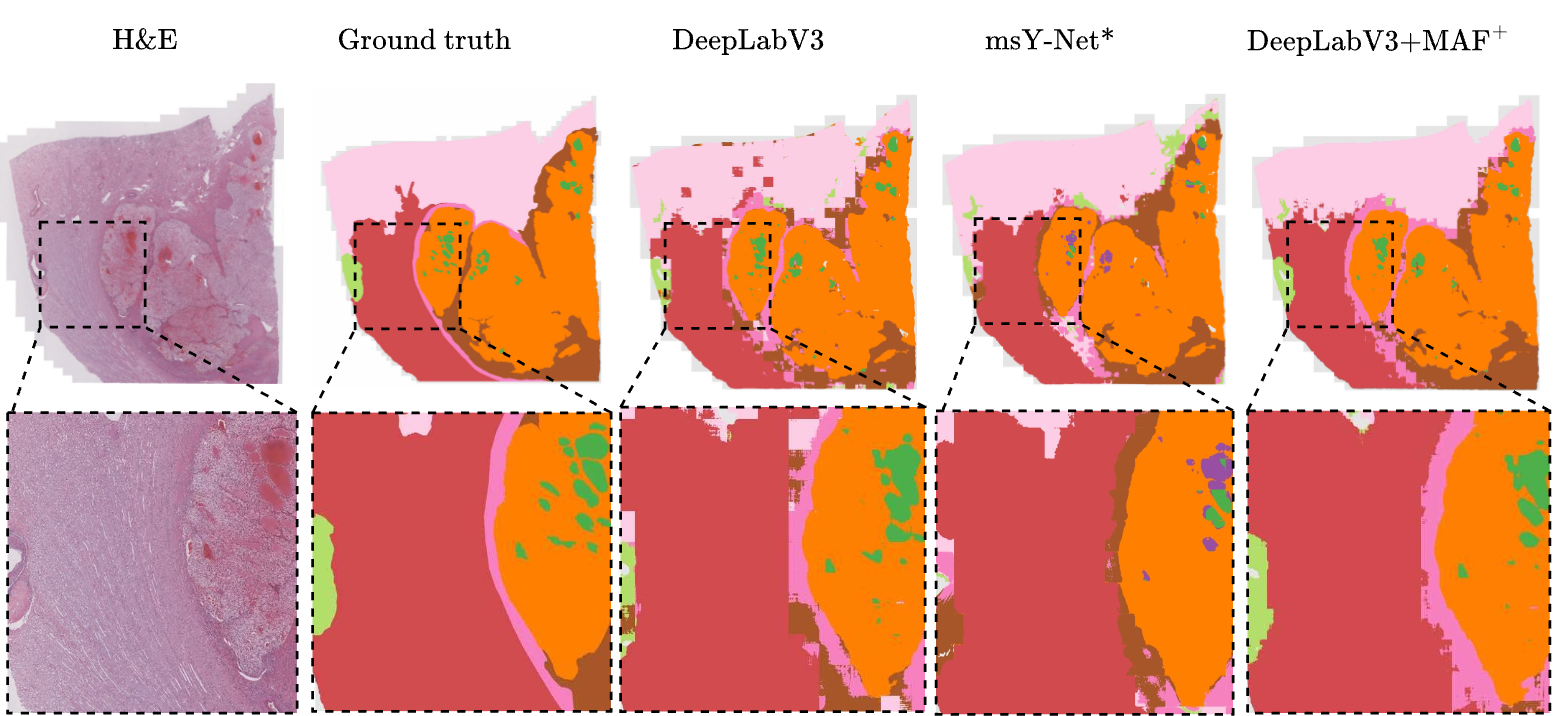}
	
	\caption{Additional comparison of RCC segmentations with $\mathit{ds}=16$ at different magnifications. *based on ResNet-18} 
	\label{add_wsi_quali}
\end{figure*}

%%%%%%%%%%%%%%%%   End   %%%%%%%%%%%%%%%%
%\end{multicols}  % Method B for two-column formatting (doesn't play well with line numbers), comment out if using method A
\end{document}